\documentclass{aero}
\usepackage{amssymb}
\usepackage{amsmath}
\usepackage{booktabs}
\usepackage{multirow}
\usepackage{graphicx}
\usepackage{subcaption}
\usepackage{float}
\usepackage[section]{placeins}
\usepackage[numbers]{natbib}
\usepackage{hyperref}
\usepackage{textcomp}
\usepackage[version=4]{mhchem}
\usepackage{siunitx}
\usepackage{makecell}
\usepackage{algpseudocode}
\usepackage{algorithm}
\usepackage{longtable}
\hypersetup{colorlinks=true,citecolor=blue,linkcolor=blue,urlcolor=blue}
\algrenewcommand{\algorithmicrequire}{\textbf{Input:}}
\algrenewcommand{\algorithmicensure}{\textbf{Output:}}

\ADsetup{
title = {Neural-Network Approximators for Low-Thrust Multi-Asteroid Rendezvous Optimization},
author = {
  Xingyu Zhou$^{1}$,
  Roberto Armellin$^{2}$,
  Jack Yarndley$^{2}$,
  and Xiangyu Li$^{3}$\cor{}
},
address = {
  1. State Key Laboratory of Lunar and Planetary Sciences, Space Science Institute, Macau University of Science and Technology, Macao SAR, People's Republic of China \sep
  2. Space Engineering Centre, University of Auckland, Auckland 1010, New Zealand \sep
  3. School of Aerospace Engineering, Beijing Institute of Technology, Beijing, 100081, China
},
email = {lixiangy@bit.edu.cn},
abstract = {
  Designing low-thrust multi-asteroid rendezvous missions requires the coupled selection of asteroid sequences, encounter epochs, and low-thrust trajectories. Lambert transfers support rapid beam search but do not directly describe thrust-limited transfer capacity or propellant consumption. This paper develops two complementary low-thrust neural approximators. The first directly predicts the continuous maximum admissible initial mass from numerically optimized low-thrust transfers, providing a transfer-capacity estimate for comparison with the current spacecraft mass. The second predicts mass-dependent propellant consumption and propagates spacecraft mass to subsequent legs. With the carefully-designed input features and network structure, the approximators reduce mean absolute errors by 72.2\% and 75.5\%, respectively, relative to conventional deep neural network baselines. These predictions are incorporated into a Lambert-ranked beam search through resource filtering and recursive mass propagation before final beam selection. The method is evaluated on fixed-visit, maximum-visit, and deployment--collection missions. Across the tested cases, the proposed resource filtering increases the fraction of candidate sequences that yield accepted low-thrust solutions, raises the maximum validated asteroid-visit count by up to 15.4\%, and improves the best mission objective by up to 17.8\%. These results demonstrate that continuous transfer-capacity and propellant predictions can improve finite-width sequence search.
},
keywords = {Low-thrust trajectory optimization \sep Multi-asteroid rendezvous \sep Neural-network approximation \sep GTOC \sep Beam search}
}

\begin{document}

\maketitle

\section{Introduction}
\label{sec1}

Multi-asteroid missions allow a single spacecraft to rendezvous with several targets within one tour, thereby increasing the scientific or operational return of a mission \cite{Zhu2009MultiAsteroid,Tang2016Adjoint}. Their design requires the target sequence, rendezvous times, and transfer trajectories to be determined together. For low-thrust spacecraft, these decisions are strongly coupled because the feasibility and propellant consumption of each transfer depend on both the available flight time and the spacecraft mass. Multi-asteroid tour design therefore combines a discrete sequence-planning problem with continuous low-thrust trajectory optimization \cite{DiCarlo2018MainBelt}. More complex mission scenarios may also include payload exchange, repeated visits, or resource collection. A representative example is the asteroid-mining problem considered in the 12th Global Trajectory Optimization Competition (GTOC12) \cite{Zhang2025GTOC12Design}.

The number of possible target sequences increases rapidly with the number of candidate asteroids and encounters. At the same time, evaluating a candidate sequence accurately requires the solution of a continuous low-thrust optimal-control problem. Numerical trajectory-optimization methods can obtain high-fidelity low-thrust solutions for a prescribed sequence and set of rendezvous times \cite{Betts1998Survey}, including sequential convex programming (SCP) methods with adaptive mesh refinement \cite{Zhao2023AdaptiveSCP}. Solving such problems repeatedly inside a large combinatorial search, however, is computationally expensive \cite{Shirazi2018Review}. For this reason, multi-target trajectory design commonly adopts a multi-fidelity strategy in which inexpensive transfer models are used during the sequence search and only a limited number of promising sequences are subsequently refined using higher-fidelity trajectory optimization. Examples include finite-Fourier-series approximations combined with tree search \cite{Fan2019MCTS} and fast sequence-selection methods for multispacecraft missions \cite{Qi2022MCTS}.

Beam search provides a practical way to control the size of the sequence search. At each search depth, candidate extensions are generated from the current partial sequences, ranked using a prescribed cost measure, and only a limited number are retained for further expansion. Lambert transfers are frequently used as the low-fidelity model because they provide a fast estimate of the cost between two prescribed orbital states \cite{Izzo2015Lambert}. Beam-based asteroid-chain construction followed by continuous-thrust optimization has been used in several GTOC solutions, including GTOC11 \cite{GTOC11Tsinghua,GTOC11Eccentric}. In GTOC12, the ACT\&Friends \cite{Izzo2025ACTFriends} and TheAntipodes \cite{TheAntipodes2025} teams also combined approximate transfer models, sequence search, and subsequent low-thrust trajectory optimization.

The main limitation of Lambert-based search is that impulsive transfer cost and low-thrust feasibility describe different properties of a transfer. A candidate with a favorable Lambert cost may still be difficult or infeasible for a thrust-limited spacecraft at its current mass and within the available transfer time. This difference becomes important in a finite-width beam search. A candidate retained at one search level occupies one of the limited beam slots and determines which branches remain available for later expansion. Therefore, a sequence that appears attractive under the Lambert model may eliminate another sequence before either one is examined with a low-thrust trajectory model. Increasing the beam width preserves more alternatives, but it does not remove the difference between the impulsive ranking model and the low-thrust dynamics. The resulting beam may therefore allocate limited survivor slots to candidates that are attractive under the impulsive model but are unlikely to admit the corresponding low-thrust transfers.

Fast approximations make large low-thrust searches tractable by avoiding a trajectory optimization for every candidate transfer. Hennes et al. developed continuous approximations of maximum initial mass and arrival mass from transfer information derived from astrodynamical analysis \cite{Hennes2016FastApproximators}. Maximum initial mass measures the largest spacecraft mass that can complete a prescribed low-thrust transfer within a fixed flight time. Neural surrogates have subsequently been used to estimate asteroid-transfer parameters, low-thrust transfer cost, fuel consumption, feasibility, and transfer time from orbital features \cite{Shang2018GPR,Zhu2019MLP,Li2020DeepSurrogates,Dempster2021OnlineDNN,Guo2023TransferTime,Ozaki2022AsteroidDNN}. Large-scale pretrained approximators have also been developed for low-thrust transfer cost and reachability across orbital environments \cite{Zhang2026PretrainedApproximators}.

These approximations have begun to enter combinatorial mission searches. The GTOC12 ACT\&Friends study combined continuous maximum-initial-mass and transfer-time approximations with neural estimates of final mass and propellant consumption during tree search \cite{Izzo2025ACTFriends}. Its neural inputs included Lambert-transfer information and relative orbital states in a departure-body frame. Recent work has also formulated low-thrust reachability through maximum initial mass and trained data-driven surrogates for that indicator \cite{Acciarini2026MIMReachability}. In its two-body electric-propulsion setting, the learned model evaluates a binary reachability indicator at a prescribed spacecraft mass rather than regressing a continuous maximum-initial-mass value.

For a multi-leg mission, the spacecraft mass changes after every transfer. Sequence search therefore requires a continuous estimate of the admissible transfer mass together with a mass-dependent estimate of propellant consumption. This study develops a Lambert-ranked beam-search method with multivariate attention neural network (MANN) resource filtering: the maximum-admissible-initial-mass prediction screens candidate extensions, and the propellant prediction propagates branch mass before final survivor allocation.
The contributions are threefold. First, two complementary low-thrust approximators are developed for direct continuous prediction of maximum admissible initial mass and mass-dependent propellant consumption. The former is trained on maximum-mass solutions obtained from low-thrust optimization and provides a continuous transfer-capacity estimate, whereas the latter predicts propellant use at the current spacecraft mass for recursive mass propagation. Second, the two predictions are incorporated into Lambert-ranked beam search as a resource filter before final beam survivor allocation. Low-thrust resource information thereby affects candidate retention while Lambert transfer information remains the ranking metric. Third, the method is evaluated on fixed-visit, maximum-visit, and deployment--collection missions with distinct objectives, visit rules, and spacecraft-mass evolution. Candidate sequences retained by the search are subsequently refined and assessed using SCP.

The remainder of this paper is organized as follows. Section~\ref{sec2} formulates the three mission classes and their common dynamics. Section~\ref{sec3} presents the neural-network approximators, dataset generation, training procedure, and prediction results. Section~\ref{sec4} describes the beam-search architectures, the use of the neural-network predictions during sequence search, and the low-thrust trajectory refinement. Section~\ref{sec5} presents the numerical results and analyzes representative mission cases. Section~\ref{sec6} concludes the paper.

\section{Multi-Asteroid Rendezvous Optimization Problem}
\label{sec2}

Multi-asteroid low-thrust mission design couples discrete target selection and visit order with continuous encounter epochs and thrust histories. This study considers three single-spacecraft mission classes that share the same dynamics but differ in their visit rules, payload changes, and objectives. The following formulation separates these common physical requirements from the task-specific constraints; the case settings are introduced with the numerical comparisons in Sec.~\ref{sec5}.

\subsection{Mission Representation}
\label{sec2.1}

Let $\mathcal{A}$ be a finite asteroid catalogue with prescribed heliocentric ephemerides $\boldsymbol{r}_i(t)$ and $\boldsymbol{v}_i(t)$ for each $i\in\mathcal{A}$. A mission selects a subset $\mathcal{A}_{\mathrm{sel}}\subseteq\mathcal{A}$ containing $N=|\mathcal{A}_{\mathrm{sel}}|$ distinct asteroids. Its encounter sequence and timing are represented by
\begin{equation}
    \boldsymbol{s}=(s_0,\ldots,s_K),\qquad
    \boldsymbol{t}=(t_0,\ldots,t_K),\qquad
    \Delta t_k=t_{k+1}-t_k,\quad k=0,\ldots,K-1.
    \label{eq1}
\end{equation}
Here, $s_k$ identifies the body encountered at $t_k$, and $s_0$ denotes Earth departure. The sequence contains $K+1$ events connected by $K$ transfer legs. The number of distinct asteroids $N$ differs from the event count when a mission includes repeated visits or Earth return.

The spacecraft position, velocity, and mass are denoted by $\boldsymbol{r}(t)$, $\boldsymbol{v}(t)$, and $m(t)$, respectively. The thrust vector $\boldsymbol{F}_k(t)$ controls leg $k$, and $m_{\mathrm{init}}$ denotes the mission launch mass. The sequence specifies the discrete choices; encounter epochs and thrust histories specify the continuous decisions. Whether the visit count and launch mass are prescribed depends on the mission class.

The mission timeline is restricted by an admissible epoch window, a maximum duration, and leg-specific transfer-time bounds:
\begin{equation}
    \begin{aligned}
        &t_{\mathrm{open}}\le t_0<t_1<\cdots<t_K\le t_{\mathrm{close}},\\
        &t_K-t_0\le T_{\mathrm{mission}},\qquad
        \Delta t_{k,\min}\le\Delta t_k\le\Delta t_{k,\max},
        \quad k=0,\ldots,K-1.
    \end{aligned}
    \label{eq2}
\end{equation}
Payload exchanges are treated as instantaneous. A coast interval between successive encounters with the same body is represented explicitly as a leg; no additional variable is introduced to represent coast duration. Section~\ref{sec2.3} specifies the admissible visit sequences.

\subsection{Low-Thrust Dynamics}
\label{sec2.2}

The spacecraft dynamics and target ephemerides are expressed in the heliocentric inertial frame under two-body gravity. Third-body perturbations and gravity-assist maneuvers are not included within this formulation. Between encounters, the spacecraft satisfies
\begin{equation}
    \dot{\boldsymbol{r}}=\boldsymbol{v},\qquad
    \dot{\boldsymbol{v}}=-\frac{\mu_\odot}{\|\boldsymbol{r}\|^3}\boldsymbol{r}
        +\frac{\boldsymbol{F}}{m},\qquad
    \dot{m}=-\frac{\|\boldsymbol{F}\|}{I_{\mathrm{sp}}g_0},
    \label{eq3}
\end{equation}
subject to the thrust limit
\begin{equation}
    \|\boldsymbol{F}(t)\|\le F_{\max}.
    \label{eq4}
\end{equation}
The propulsion model assumes constant specific impulse $I_{\mathrm{sp}}$ and maximum thrust $F_{\max}$, while the thrust magnitude and direction may vary.

An asteroid rendezvous requires both the position and velocity to match the target:
\begin{equation}
    \boldsymbol{r}(t_k)=\boldsymbol{r}_{s_k}(t_k),\qquad
    \boldsymbol{v}(t_k)=\boldsymbol{v}_{s_k}(t_k),
    \qquad s_k\in\mathcal{A}.
    \label{eq5}
\end{equation}
At Earth departure and Earth return, the position must match Earth's position, while hyperbolic excess speed is bounded above by $v_{\infty,\max}$:
\begin{equation}
    \boldsymbol{r}(t_e)=\boldsymbol{r}_{\mathrm{E}}(t_e),\qquad
    \|\boldsymbol{v}(t_e)-\boldsymbol{v}_{\mathrm{E}}(t_e)\|
        \le v_{\infty,\max}.
    \label{eq6}
\end{equation}
Here, $t_e$ is the departure or return epoch, and $v_{\infty,\max}$ is the applicable mission limit.

Position and velocity remain continuous through each encounter, but payload exchange may change the spacecraft mass instantaneously. Let $m_k^-$ and $m_k^+$ denote its values immediately before and after event $k$. The mass balances are
\begin{equation}
    \begin{aligned}
        m_{k+1}^-&=m_k^+-
            \int_{t_k}^{t_{k+1}}
            \frac{\|\boldsymbol{F}_k(t)\|}{I_{\mathrm{sp}}g_0}\,\mathrm{d}t,
            && k=0,\ldots,K-1,\\
        m_k^+&=m_k^-+\Delta m_k^{\mathrm{event}},
            && k=1,\ldots,K.
    \end{aligned}
    \label{eq7}
\end{equation}
The first relation accounts for propellant consumption; the second accounts for event-related mass changes. The launch state satisfies $m_0^+=m_{\mathrm{init}}\le m_{\mathrm{init,max}}$, where $m_{\mathrm{init,max}}$ is the launch-mass limit. All trajectories must satisfy $m(t)\ge m_{\mathrm{dry}}$; payload-dependent lower bounds and the event mass changes are specified by the mission class.

\subsection{Mission Classes}
\label{sec2.3}

The three mission classes specialize the common model through their sequence restrictions, launch-mass conditions, and payload rules. Define the terminal spacecraft mass as $m_f=m_K^-$, before any terminal payload unloading.

\subsubsection{Fixed-Visit Rendezvous}
\label{sec2.3.1}

A fixed-visit tour starts at Earth, visits a prescribed number $N$ of distinct asteroids, and terminates at the last asteroid. The design variables are the selected targets, their visit order, the encounter epochs, and the thrust histories; the launch mass is prescribed. The subsequence $(s_1,\ldots,s_N)$ must be a permutation of $\mathcal{A}_{\mathrm{sel}}$, so each selected asteroid is visited once and $K=N$.

No payload is deployed or collected, giving $\Delta m_k^{\mathrm{event}}=0$ and the mass lower bound $m_{\mathrm{dry}}$. The objective is to maximize $m_f$ subject to the common constraints. With fixed launch mass and no event mass exchange, this is equivalent to minimizing total propellant consumption.

\subsubsection{Maximum-Visit Rendezvous}
\label{sec2.3.2}

The maximum-visit problem retains the single-visit rule, fixed launch mass, and zero event mass changes, but does not prescribe $N$. The design therefore includes the sequence length as well as target selection, visit order, encounter epochs, and thrust histories. The visit count follows from the selected sequence, $N=|\mathcal{A}_{\mathrm{sel}}|$, rather than constituting an independent choice.

The objective is to maximize $(N,m_f)$ in lexicographic order. A feasible tour with more visited asteroids takes priority over one with fewer visits, irrespective of their terminal spacecraft masses. Terminal spacecraft mass distinguishes feasible tours only when their visit counts are equal.

\subsubsection{Deployment--Collection}
\label{sec2.3.3}

A deployment--collection mission visits each of $N$ selected asteroids to deploy a miner, revisits it to collect mined material, and returns to Earth. The mission class considered here prescribes $N$ and requires all deployments to precede all collections. The deployment subsequence $(s_1,\ldots,s_N)$ and collection subsequence $(s_{N+1},\ldots,s_{2N})$ are permutations of the same set $\mathcal{A}_{\mathrm{sel}}$. Their orders may differ, and the final Earth event gives $K=2N+1$.

The design variables are the selected targets, both visit orders, the encounter epochs, and the thrust histories. Launch mass is additionally optimized when only an upper bound is prescribed; otherwise, it is fixed. The launch mass includes the dry spacecraft, propellant, and the miners carried for subsequent deployment.

Let $M_i$ denote the mined mass collected from asteroid $i$ and $M_{\mathrm{ret}}=\sum_{i\in\mathcal{A}_{\mathrm{sel}}}M_i$ the returned mined mass. Each deployment releases one miner of mass $m_{\mathrm{miner}}$, whereas each collection adds mined material without recovering the miner. For events $k=1,\ldots,K$, the mass changes are
\begin{equation}
    \Delta m_k^{\mathrm{event}}=
    \begin{cases}
        -m_{\mathrm{miner}}, & \text{miner deployment},\\
        M_{s_k}, & \text{mined-material collection},\\
        -M_{\mathrm{ret}}, & \text{terminal Earth unloading}.
    \end{cases}
    \label{eq8}
\end{equation}
The miner operates while the spacecraft continues its tour, so mining duration is distinct from spacecraft dwell time. Under constant-rate mining and full collection,
\begin{equation}
    M_i=\dot{M}_{\mathrm{mine}}
        \bigl(t_i^{\mathrm{col}}-t_i^{\mathrm{dep}}\bigr),\qquad
    t_i^{\mathrm{col}}-t_i^{\mathrm{dep}}\ge\tau_{\mathrm{mine,min}},
    \qquad i\in\mathcal{A}_{\mathrm{sel}}.
    \label{eq9}
\end{equation}
Here, $t_i^{\mathrm{dep}}$ and $t_i^{\mathrm{col}}$ are the deployment and collection epochs. The parameters $\dot{M}_{\mathrm{mine}}$ and $\tau_{\mathrm{mine,min}}$ specify the mining rate and minimum operating interval.

To maintain nonnegative propellant mass, the spacecraft mass must satisfy
\begin{equation}
    m(t)\ge m_{\mathrm{dry}}
        +n_{\mathrm{rem}}(t)m_{\mathrm{miner}}+M_{\mathrm{car}}(t).
    \label{eq10}
\end{equation}
The quantity $n_{\mathrm{rem}}(t)$ counts miners still onboard, and $M_{\mathrm{car}}(t)$ is the onboard mined mass. This bound applies throughout each transfer and to the corresponding one-sided values before and after every mass-changing event. Immediately before Earth unloading, it becomes $m_K^-\ge m_{\mathrm{dry}}+M_{\mathrm{ret}}$.

The objective is to maximize $M_{\mathrm{ret}}$, excluding the spacecraft and remaining propellant. Encounter timing affects both the amount mined and the payload carried on subsequent transfers, coupling the mining objective to low-thrust feasibility. Table~\ref{tab1} summarizes the three objectives; their benchmark implementations are described in Sec.~\ref{sec5}.

\begin{table}[!h]
    \caption{Visit counts and objectives of the three mission classes.}
    \label{tab1}
    \centering
    \begin{tabular}{@{}lll@{}}
        \hline\hline
        Mission class & Visit count & Objective \\
        \midrule
        Fixed-visit rendezvous & Prescribed $N$ & $\max m_f$ \\
        Maximum-visit rendezvous & Variable $N$ & $\max_{\mathrm{lex}}(N,m_f)$ \\
        Deployment--collection & Prescribed $N$ & $\max M_{\mathrm{ret}}$ \\
        \hline\hline
    \end{tabular}
\end{table}

\subsection{Problem Formulation}
\label{sec2.4}

Let $\mathcal{S}$ denote the set of encounter sequences satisfying the visit rules of the selected mission class in Sec.~\ref{sec2.3}. With the corresponding objective $J$ from Table~\ref{tab1}, the coupled problem is
\begin{equation}
    \begin{aligned}
        \underset{\boldsymbol{s},\,\boldsymbol{t},\,m_{\mathrm{init}},\,
            \{\boldsymbol{F}_k(\cdot)\}_{k=0}^{K-1}}{\operatorname{maximize}}
        \quad & J,\\
        \text{subject to}\quad
        &\boldsymbol{s}\in\mathcal{S},\\
        &\text{time bounds in Eq.~\eqref{eq2}},\\
        &\text{dynamics and thrust limit in Eqs.~\eqref{eq3}--\eqref{eq4}},\\
        &\text{applicable encounter conditions in Eqs.~\eqref{eq5}--\eqref{eq6}},\\
        &\text{mass balances in Eq.~\eqref{eq7}},\\
        &m_0^+=m_{\mathrm{init}}\le m_{\mathrm{init,max}},\quad m(t)\ge m_{\mathrm{dry}},\\
        &\text{task-specific mass conditions in Sec.~\ref{sec2.3}}.
    \end{aligned}
    \label{eq11}
\end{equation}
For the two rendezvous classes, the task-specific mass conditions fix $m_{\mathrm{init}}$ and set $\Delta m_k^{\mathrm{event}}=0$. For deployment--collection, they impose Eqs.~\eqref{eq8}--\eqref{eq10} and either a prescribed launch mass or an upper-bound condition. The launch mass is a decision variable only under the upper-bound condition. The coupled sequence, timing, and mass requirements motivate the transfer approximators in Sec.~\ref{sec3} and the search and trajectory-refinement procedure in Sec.~\ref{sec4}.

\section{Neural-Network-Based Low-Thrust Approximation}
\label{sec3}

Two neural networks estimate complementary transfer quantities: the maximum admissible initial mass at prescribed boundary states and duration, and the propellant consumption at a specified initial mass. The maximum-admissible-initial-mass (MIM) model defines the transfer-capacity boundary and, during search, provides the capacity-related features used by the propellant model. It also predicts the terminal mass of the maximum-initial-mass solution; the propellant model predicts the terminal mass at the prescribed initial mass. Both models provide output-dependent standard deviations for surrogate-error diagnostics. This section defines the approximators and evaluates their single-leg predictions; Sec.~\ref{sec4} describes their use in sequence search.

\subsection{Neural Regression Framework}
\label{sec3.1}

Both low-thrust approximators employ the MANN introduced by Zhang et al.~\cite{Zhang2024MANN} for Earth--Jupiter orbit-deviation propagation. In that application, MANN showed higher prediction accuracy and better generalization than a fully connected deep neural network. Here, MANN is the common regression backbone for the MIM and propellant models. The two models are trained separately and evaluated against fully connected deep neural networks (DNNs) and Kolmogorov--Arnold networks (KANs) on matched low-thrust datasets and test conditions.

For a normalized input $\widetilde{\boldsymbol{x}}$, each model returns two normalized means and two log variances:
\begin{equation}
    f_{\boldsymbol{\theta}}(\widetilde{\boldsymbol{x}})
    =(\widetilde{\mu}_1,\widetilde{\mu}_2,\ell_1,\ell_2),
    \qquad
    \widetilde{\sigma}_j^2=\exp(\ell_j),\quad j=1,2,
    \label{eq12}
\end{equation}
Here, $\boldsymbol{\theta}$ collects the trainable parameters. Assuming conditionally independent Gaussian outputs, the Gaussian negative log-likelihood (GNLL) is averaged over $n_{\mathrm{batch}}$ samples and $n_y=2$ outputs:
\begin{equation}
    \mathcal{L}_{\mathrm{GNLL}}
    =\frac{1}{2n_{\mathrm{batch}}n_y}
      \sum_{b=1}^{n_{\mathrm{batch}}}\sum_{j=1}^{n_y}
      \left[
        \ell_{bj}
        +\frac{(\widetilde{y}_{bj}-\widetilde{\mu}_{bj})^2}
        {\exp(\ell_{bj})+\varepsilon_{\mathrm{num}}}
      \right].
    \label{eq13}
\end{equation}
The additive constant is omitted, and $\varepsilon_{\mathrm{num}}$ is a numerical regularizer. This heteroscedastic formulation estimates the conditional mean and standard deviation of each output \cite{Nix1994MeanVariance}. Inputs and labels are scaled componentwise to the unit interval. Mean absolute error (MAE) and mean relative error (MRE) are evaluated after inverse scaling.

Training progressively transitions from mean-square error (MSE) to GNLL:
\begin{equation}
    \mathcal{L}_e=(1-w_e)\mathcal{L}_{\mathrm{MSE}}
                    +w_e\mathcal{L}_{\mathrm{GNLL}},
    \qquad
    w_e=\min\!\left(\frac{e}{E_{\mathrm{w}}},1\right),
    \label{eq14}
\end{equation}
where $e$ is the zero-based epoch index and $E_{\mathrm{w}}$ is the transition length. MSE is averaged over samples and outputs. The transition first emphasizes the mean response, then jointly fits means and variances. The predicted means are used for resource filtering and recursive mass propagation, whereas the predicted standard deviations are retained for surrogate-error diagnostics.

Table~\ref{tab2} lists the selected configurations and shared optimization settings. The development sets contain 50,000, 100,000, 250,000, or one million accepted records, and each task has an independently generated 100,000-record test set. Input and output scaling is determined from the development data and applied unchanged to the test set. The selected million-record models use a 95/5 training--validation split. Multiple propellant records derived from one transfer may appear in both development subsets, whereas the test conditions are disjoint from the development data. Model configurations and checkpoints are selected by validation GNLL after $w_e=1$ in Eq.~\eqref{eq14}; all reported errors use the independent test sets.

For the 250,000-record model-family comparison, DNN, KAN, and MANN use the same task-specific development records, row-level training--validation partition, normalization bounds, and independently generated 100,000-record test set. Within each model family, the reported architecture and checkpoint are selected by validation GNLL from the evaluated configuration grid, and the independent test set is reserved for final performance evaluation.

\begin{table}[!h]
\caption{Selected network configurations and training settings.}
\label{tab2}
\centering
\begin{tabular}{@{}lll@{}}
\hline\hline
Setting & MIM & Propellant\\
\midrule
Input dimension & 27 & 29\\
Latent dimension / head dimension & 64 / 64 & 64 / 64\\
Blocks / attention heads per block & 6 / 6 & 10 / 10\\
GNLL transition $E_{\mathrm{w}}$ & 200 epochs & 100 epochs\\
Predicted-feature transition & --- & Epochs 100--500\\
\midrule
Optimizer & \multicolumn{2}{l}{Adam}\\
Initial learning rate / weight decay & \multicolumn{2}{l}{$10^{-3}$ / $10^{-5}$}\\
Learning-rate multiplier & \multicolumn{2}{l}{0.5 every 100 epochs}\\
Epoch budget / maximum batch size & \multicolumn{2}{l}{1000 / 50,000}\\
Random seed & \multicolumn{2}{l}{42}\\
\hline\hline
\end{tabular}
\end{table}

\subsection{Maximum Admissible Initial-Mass Prediction}
\label{sec3.2}

\subsubsection{Prediction Target and Representation}
\label{sec3.2.1}

Consider a transfer of duration $T$ between Cartesian boundary states $\boldsymbol{\xi}_a=(\boldsymbol{r}_a,\boldsymbol{v}_a)$ and $\boldsymbol{\xi}_b=(\boldsymbol{r}_b,\boldsymbol{v}_b)$. The single-leg initial and terminal masses are $m_a$ and $m_b$. The maximum admissible initial mass satisfies
\begin{equation}
    \begin{aligned}
        m_{a,\max}(\boldsymbol{\xi}_a,\boldsymbol{\xi}_b,T)
        =\underset{m_a,\,\boldsymbol{F}(\cdot)}{\operatorname{maximize}}
        \quad &m_a,\\
        \text{subject to}\quad
        &\text{dynamics and thrust limit in Eqs.~\eqref{eq3}--\eqref{eq4}},\\
        &(\boldsymbol{r}(0),\boldsymbol{v}(0))=\boldsymbol{\xi}_a,\\
        &(\boldsymbol{r}(T),\boldsymbol{v}(T))=\boldsymbol{\xi}_b,\\
        &m(0)=m_a,\qquad m(T)\ge m_{\mathrm{lb}} .
    \end{aligned}
    \label{eq15}
\end{equation}
The labels are $\boldsymbol{y}_{\mathrm{M}}=(m_{a,\max},m_{b,\max})$, obtained from the same numerically optimized transfer. The second component denotes the terminal mass associated with maximum initial mass, not a separately maximized terminal mass. The transfer contains no payload exchange, and $m_{\mathrm{lb}}$ is its terminal mass floor. Unlike fixed-mass binary reachability prediction, the MIM model used here regresses the continuous capacity boundary $m_{a,\max}$ for the prescribed boundary states and transfer duration. The current spacecraft mass is not an input to the MIM model; for fixed transfer conditions, different spacecraft masses can therefore be evaluated against the same predicted boundary.

The rotational symmetry of the solar two-body dynamics permits a common coordinate representation for differently oriented transfers. For a departure state with nonzero angular momentum, define the orthonormal basis and rotation
\begin{equation}
    \boldsymbol{e}_r=\frac{\boldsymbol{r}_a}{\|\boldsymbol{r}_a\|},\qquad
    \boldsymbol{e}_h=\frac{\boldsymbol{r}_a\times\boldsymbol{v}_a}
                          {\|\boldsymbol{r}_a\times\boldsymbol{v}_a\|},\qquad
    \boldsymbol{e}_t=\boldsymbol{e}_h\times\boldsymbol{e}_r,\qquad
    \boldsymbol{R}=\left[\boldsymbol{e}_r\ \boldsymbol{e}_t\ \boldsymbol{e}_h\right]^{\mathsf T}.
    \label{eq16}
\end{equation}
The same rotation is applied to both boundary states and held fixed throughout the transfer. In this canonical frame,
\begin{equation}
    \begin{aligned}
        \boldsymbol{r}'_j&=\boldsymbol{R}\boldsymbol{r}_j,\qquad
        \boldsymbol{v}'_j=\boldsymbol{R}\boldsymbol{v}_j,\quad j=a,b,\\
        \boldsymbol{r}'_a&=(\|\boldsymbol{r}_a\|,0,0)^{\mathsf T},\qquad
        \boldsymbol{v}'_a=(v_{r,a},v_{t,a},0)^{\mathsf T},
    \end{aligned}
    \label{eq17}
\end{equation}
where $v_{r,a}=\boldsymbol{e}_r^{\mathsf T}\boldsymbol{v}_a$ and $v_{t,a}=\boldsymbol{e}_t^{\mathsf T}\boldsymbol{v}_a$. Applying $\boldsymbol{F}'=\boldsymbol{R}\boldsymbol{F}$ preserves Eqs.~\eqref{eq3}--\eqref{eq4}, the rendezvous conditions, and the mass history, so the maximum-mass and propellant labels are unchanged. The transformation removes the overall orientation without imposing coplanarity: the initial state has three independent components, whereas all six terminal-state components are retained.

The selected representation expresses these rotated states in spherical coordinates. Let $\boldsymbol{\chi}\in\mathbb{R}^{23}$ contain the three nonconstant initial-state components, six terminal-state components, two six-component two-body propagation residuals, $T$, and the phase $T/P_a$, where $P_a$ is the initial orbital period. The residuals compare forward- and backward-propagated two-body states with the opposite boundary states. Together with the Lambert velocity increments and analytical mass features, these quantities form the input
\begin{equation}
    \boldsymbol{x}_{\mathrm{M}}
    =\bigl(\boldsymbol{\chi},\Delta v_1,\Delta v_2,
                 m_a^{\mathrm{g}},m_b^{\mathrm{g}}\bigr)
       \in\mathbb{R}^{27}.
    \label{eq18}
\end{equation}
The features depend on transfer boundary states and propulsion parameters rather than asteroid identifiers. The approximators can therefore be applied to other targets whose transfer conditions lie within the sampled domain in Table~\ref{tab3}.
The auxiliary impulsive velocity increments $\Delta v_1$ and $\Delta v_2$ define $\Delta v=\Delta v_1+\Delta v_2$ and the analytical features
\begin{equation}
    \Delta m_{\mathrm{F}}=\frac{F_{\max}T}{I_{\mathrm{sp}}g_0},
    \qquad
    m_a^{\mathrm{g}}
      =\frac{\Delta m_{\mathrm{F}}}
        {1-\exp[-\Delta v/(I_{\mathrm{sp}}g_0)]},
    \qquad
    m_b^{\mathrm{g}}=m_a^{\mathrm{g}}-\Delta m_{\mathrm{F}}.
    \label{eq19}
\end{equation}
Here, $\Delta m_{\mathrm{F}}$ is the mass consumed by continuous full thrust over $T$. Combining this quantity with the impulsive mass ratio gives analytical features, whereas Eq.~\eqref{eq15} supplies the supervised labels. The thrust limit also gives $m_a-m_b\leq\Delta m_{\mathrm{F}}$, with equality for full-thrust operation throughout the transfer. Approximating the capacity solution by this limiting mass loss yields a second initial-mass estimate, $\widehat m_{b,\max}+\Delta m_{\mathrm{F}}$. Its equal-weight average with the direct estimate defines
\begin{equation}
    \widehat{m}_{a,\max}^{\,\mathrm{fus}}
       =\frac{\widehat{m}_{a,\max}
              +\widehat{m}_{b,\max}+\Delta m_{\mathrm{F}}}{2}.
    \label{eq20}
\end{equation}
Equation~\eqref{eq20} combines two estimates of the same capacity without fitting an additional fusion weight. The terminal-mass reconstruction is exact under full-thrust operation and approximate when the optimized transfer contains reduced-thrust intervals. Section~\ref{sec3.2.4} evaluates its effect by comparing direct and fused predictions on the same samples.

\subsubsection{Training Data}
\label{sec3.2.2}

Algorithm~\ref{alg1} summarizes the common sample construction. Latin-hypercube sampling selects initial orbital conditions and transfer durations in the canonical frame. Two three-dimensional velocity increments and two-body propagation then define terminal states within the orbital domain of Table~\ref{tab3}. Each admissible transfer is labeled by solving the corresponding low-thrust optimization problem, and the boundary states are expressed in the same departure-aligned frame when the trained models are evaluated.

The sampling domain covers the central main-belt regime containing the benchmark targets in Appendix~\ref{sec7.3}; the terminal-inclination bound is measured relative to the departure plane. A label is retained when SCP converges and the mass bounds in Table~\ref{tab3} are satisfied. The 10-kg terminal floor constrains the optimization, whereas the 500-kg threshold defines the retained dataset. The GTOC12 catalogue supplies the benchmark orbits, while the supervised labels are generated independently from sampled transfer conditions.

\begin{algorithm}[!h]
\caption{Generation of single-leg training samples}
\label{alg1}
\begin{algorithmic}[1]
\Require Prediction task, sampling domain, record budget, and label-acceptance criteria.
\Ensure Accepted input--label pairs $\mathcal{D}$.
\State $\mathcal{D}\gets\varnothing$; generate canonical-frame samples $\Theta$ by Latin hypercube.
\For{each sample $\theta\in\Theta$}
    \State Construct the terminal state; \textbf{continue} if domain checks fail.
    \State Solve the maximum-initial-mass problem. \Comment{Eq.~\eqref{eq15}}
    \State \textbf{continue} if label-acceptance checks fail.
    \If{mass prediction}
        \State $\mathcal{L}\gets\{(m_{a,\max},m_{b,\max})\}$.
    \Else
        \State Sample interior initial masses. \Comment{Sec.~\ref{sec3.3.2}}
        \State Solve the propellant problems at these masses. \Comment{Eq.~\eqref{eq21}}
        \State $\mathcal{L}\gets$ boundary record and accepted interior records.
    \EndIf
    \State Construct $(\boldsymbol{x}_r,\boldsymbol{y}_r)$ for each $r\in\mathcal{L}$.
    \State $\mathcal{D}\gets\mathcal{D}\cup\{(\boldsymbol{x}_r,\boldsymbol{y}_r):r\in\mathcal{L}\}$; \textbf{break} if the budget is met.
\EndFor
\State \Return $\mathcal{D}$.
\end{algorithmic}
\end{algorithm}

\begin{table}[!h]
\caption{Sampling domain and retained datasets for both approximators.}
\label{tab3}
\centering
\begin{tabular}{@{}ll@{}}
\hline\hline
Quantity & Range or value\\
\midrule
Initial and terminal semimajor axis & 2.5--3.0 AU\\
Initial and terminal eccentricity & 0--0.4\\
Initial inclination in the canonical frame & $0^\circ$\\
Terminal inclination relative to the initial plane & $0$--$20^\circ$\\
Each velocity-increment magnitude & 0.1--2.5 km/s\\
Transfer duration & 50--500 days\\
Maximum thrust / specific impulse & 0.6 N / 4000 s\\
SCP terminal mass floor & 10 kg\\
Retained maximum initial mass & $\le20{,}000$ kg\\
Retained terminal mass & $\ge500$ kg\\
Development records & 50k, 100k, 250k, 1M\\
Test records & 100k\\
\hline\hline
\end{tabular}
\end{table}

Throughout this section, MANN configurations are denoted by $L/w$, where $L$ is the number of stacked blocks and $w$ is the latent width; the evaluated MANN models use $H=L$ attention heads per block and $d=d_h=w$, where $d$ and $d_h$ denote the latent and head dimensions, respectively.

\subsubsection{Prediction Performance}
\label{sec3.2.4}

\begin{figure}[!h]
	\centering
	\begin{subfigure}[t]{0.45\linewidth}
		\centering
		\includegraphics[width=\linewidth]{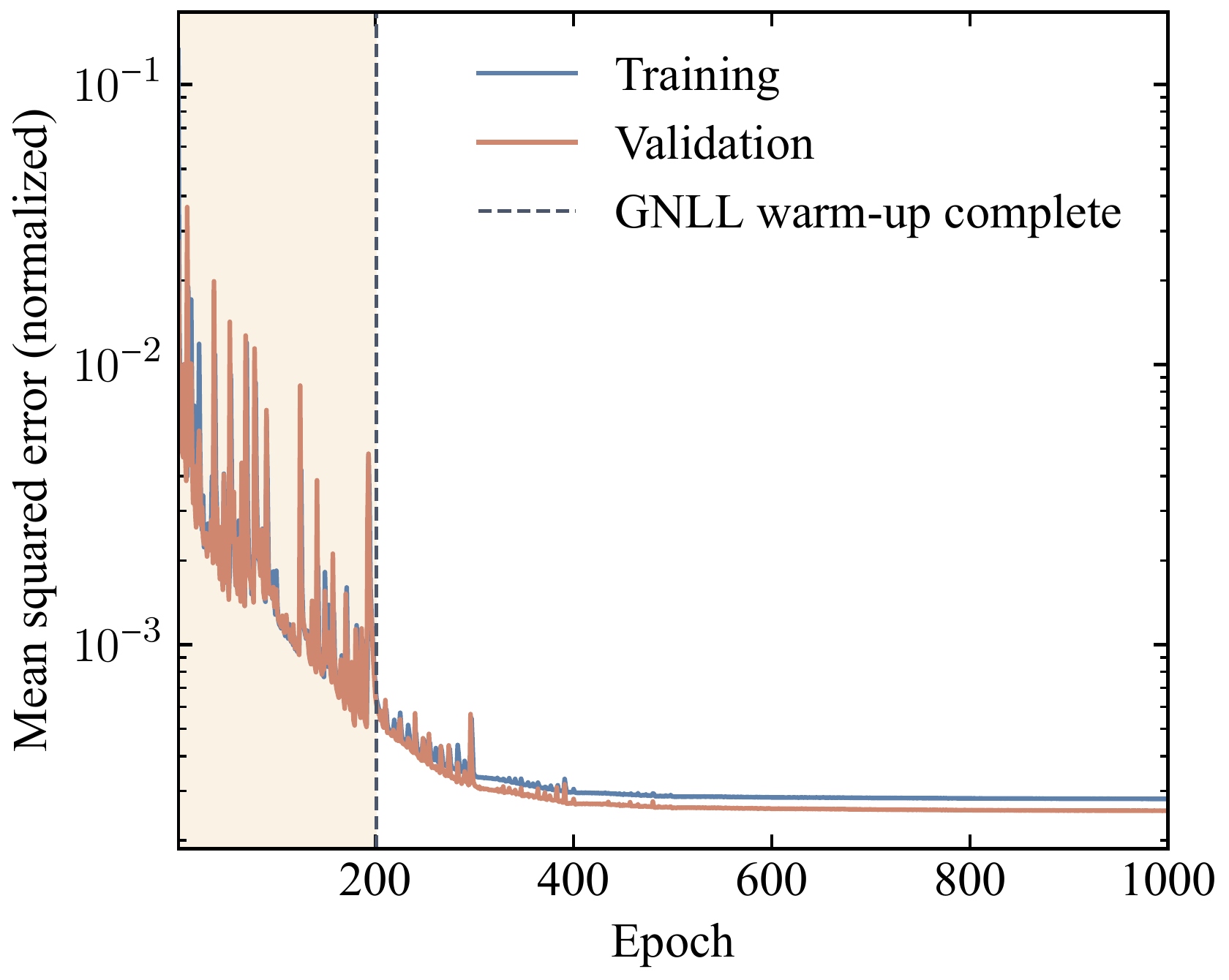}
		\caption{Training convergence.}
		\label{fig1.1}
	\end{subfigure}\hfill
	\begin{subfigure}[t]{0.45\linewidth}
		\centering
		\includegraphics[width=\linewidth]{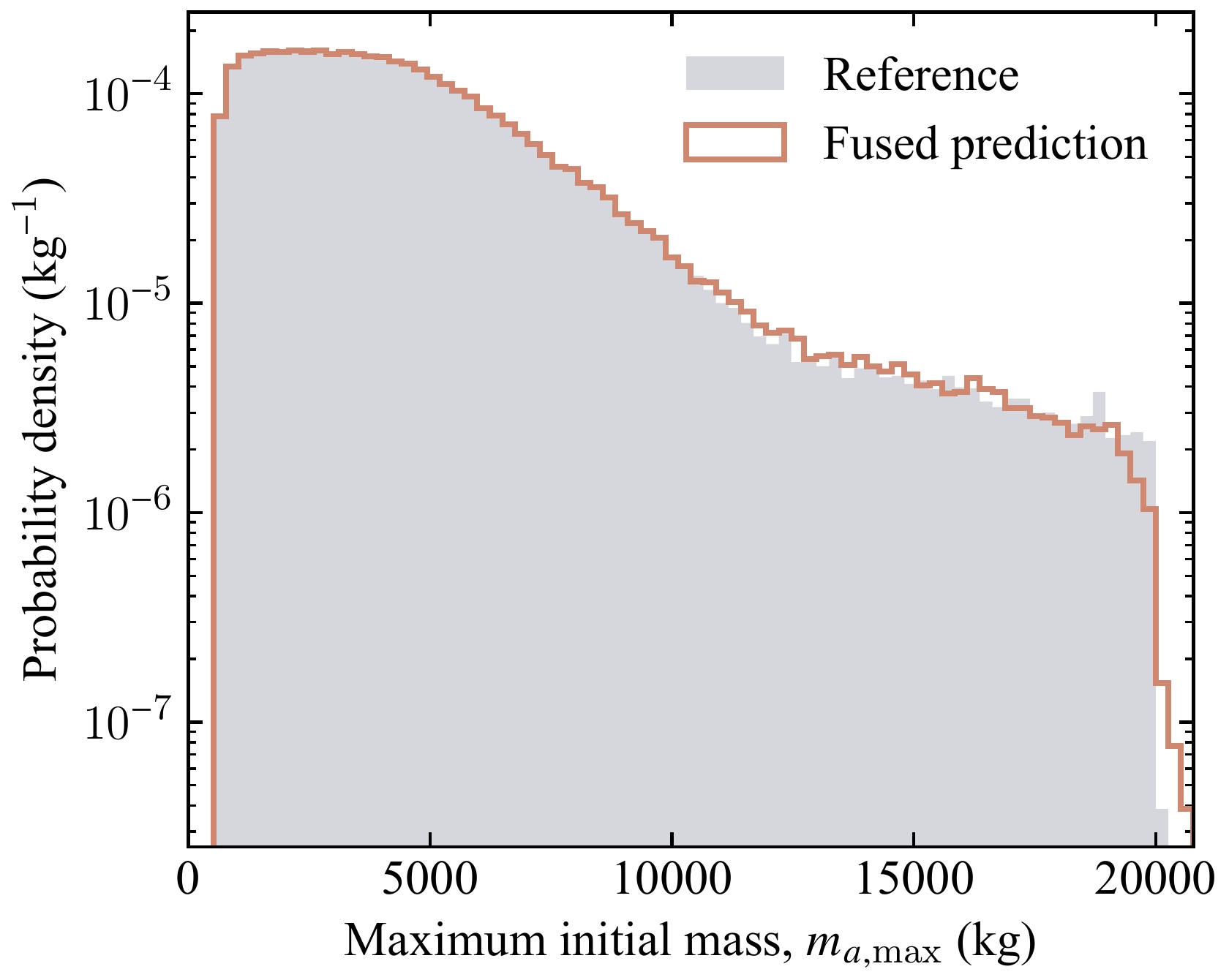}
		\caption{Mass distribution.}
		\label{fig1.2}
	\end{subfigure}
	\begin{subfigure}[t]{0.45\linewidth}
		\centering
		\includegraphics[width=\linewidth]{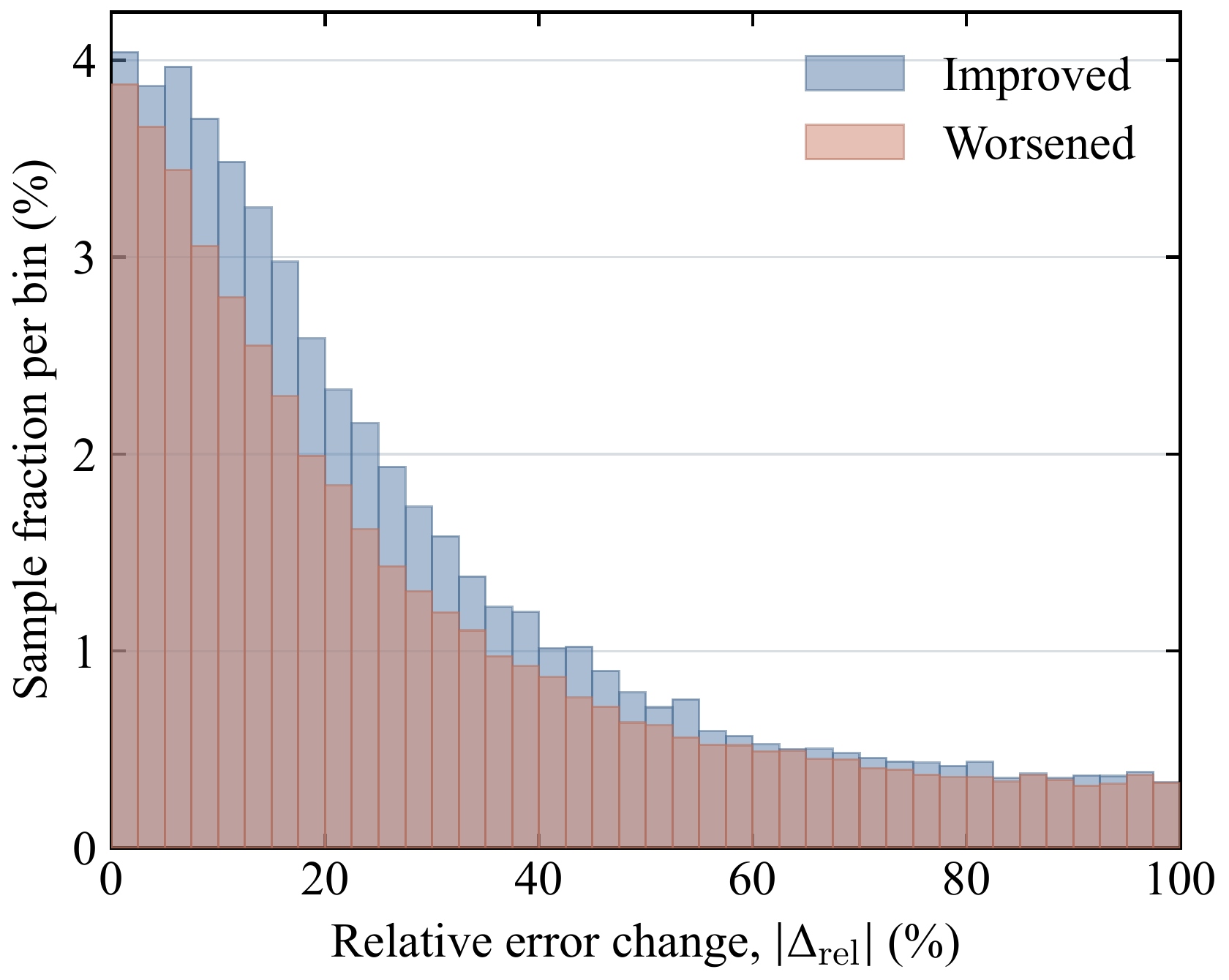}
		\caption{Effect of fusion.}
		\label{fig1.3}
	\end{subfigure}\hfill
	\begin{subfigure}[t]{0.45\linewidth}
		\centering
		\includegraphics[width=\linewidth]{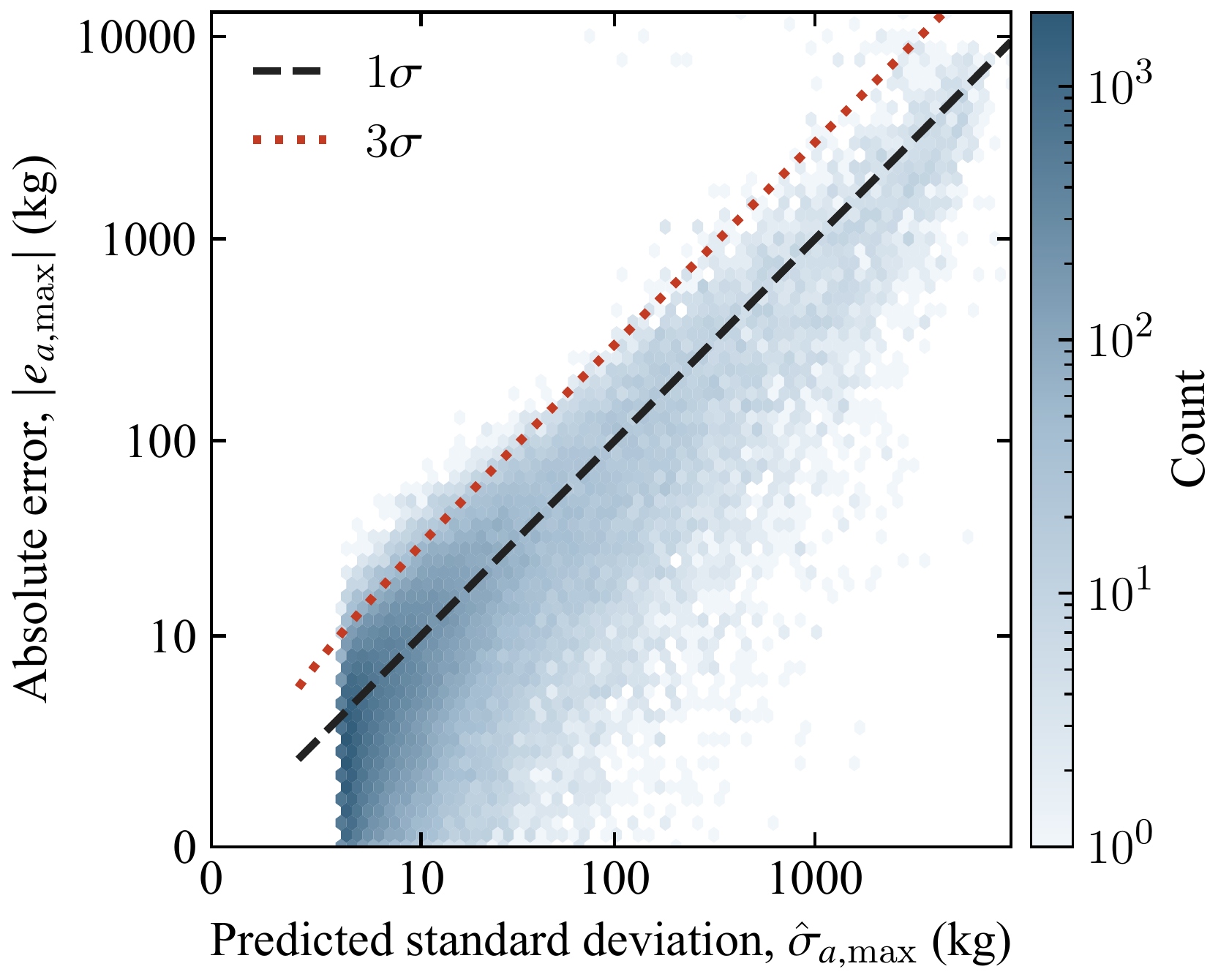}
		\caption{Surrogate-error diagnostic.}
		\label{fig1.4}
	\end{subfigure}
\caption{Training and prediction performance of the MIM MANN.}
	\label{fig1}
\end{figure}

Under the matched 250,000-record protocol, the validation-selected $6/64$ MANN yields initial- and terminal-mass MAEs of 62.88 and 63.28~kg, compared with 226.15 and 226.70~kg for DNN and 262.45 and 262.25~kg for KAN (Table~\ref{tab4}). Physically informed input features account for most of the capacity-prediction improvement: augmenting the basic state--time representation with transfer and propagation descriptors reduces the initial-mass MAE from 993.36 to 62.88~kg (Table~\ref{tabA1}). The selected model converges under staged training [Fig.~\ref{fig1}(\subref{fig1.1})]. Increasing the MANN development set to one million records further reduces the errors to 41.45 and 40.97~kg, providing a larger gain than the tested changes in MANN blocks and width.

The fused estimates reproduce the reference mass distribution and its upper tail across the 100,000 test transfers [Fig.~\ref{fig1}(\subref{fig1.2})]. Equation~\eqref{eq20} slightly reduces the initial-mass MAE from 41.45 to 40.92~kg and lowers the absolute error for 54.5\% of the samples [Fig.~\ref{fig1}(\subref{fig1.3})].

As a surrogate-error diagnostic, larger predicted standard deviations generally accompany larger direct initial-mass errors [Fig.~\ref{fig1}(\subref{fig1.4})]. The corresponding empirical coverage is 71.60, 94.85, and 99.13\% at one, two, and three predicted standard deviations, respectively.

\subsection{Mass-Dependent Propellant-Consumption Prediction}
\label{sec3.3}

\subsubsection{Prediction Target and Capacity-Related Features}
\label{sec3.3.1}

The propellant model uses the coordinate rotation and orbital descriptors of Sec.~\ref{sec3.2.1}, augmented by initial mass and capacity-related features. For prescribed boundary states, duration, and initial mass $m_a$, its propellant label satisfies
\begin{equation}
    \begin{aligned}
        \Delta m_{\mathrm{p}}^\star
          (\boldsymbol{\xi}_a,\boldsymbol{\xi}_b,T,m_a)
        =\underset{\boldsymbol{F}(\cdot)}{\operatorname{minimize}}
        \quad &m_a-m(T),\\
        \text{subject to}\quad
        &\text{dynamics and thrust limit in Eqs.~\eqref{eq3}--\eqref{eq4}},\\
        &(\boldsymbol{r}(0),\boldsymbol{v}(0))=\boldsymbol{\xi}_a,\\
        &(\boldsymbol{r}(T),\boldsymbol{v}(T))=\boldsymbol{\xi}_b,\\
        &m(0)=m_a,\qquad m(T)\ge m_{\mathrm{lb}} .
    \end{aligned}
    \label{eq21}
\end{equation}
The outputs are $\boldsymbol{y}_{\mathrm{P}}=(\Delta m_{\mathrm{p}}^\star,m_b^\star)$, where $m_b^\star=m_a-\Delta m_{\mathrm{p}}^\star$. The selected input combines the orbital descriptors with the actual initial mass, an analytical terminal-mass estimate, and two capacity-related features:
\begin{equation}
    \begin{aligned}
        \boldsymbol{x}_{\mathrm{P}}
          &=\bigl(\boldsymbol{\chi},\Delta v_1,\Delta v_2,
                    m_a,m_b^{\mathrm{L}},\rho,q\bigr)
                \in\mathbb{R}^{29},\\
        m_b^{\mathrm{L}}&=m_a
             \exp[-\Delta v/(I_{\mathrm{sp}}g_0)] .
    \end{aligned}
    \label{eq22}
\end{equation}
Here, $m_b^{\mathrm L}$ is the impulsive terminal-mass estimate at the prescribed $m_a$, distinct from the capacity estimate $m_b^{\mathrm g}$ in Eq.~\eqref{eq19}. The optimization-derived features are $\rho=m_a/m_{a,\max}$ and $q=m_{a,\max}-m_{b,\max}$. At inference they are replaced by $\widehat{\rho}=m_a/\widehat{m}_{a,\max}^{\,\mathrm{fus}}$ and $\widehat{q}=\Delta m_{\mathrm F}$. The search propagates mass using the direct propellant output.

\subsubsection{Training Data and Feature Consistency}
\label{sec3.3.2}

The propellant branch of Algorithm~\ref{alg1} extends each accepted maximum-mass solution to smaller initial masses while holding the transfer conditions fixed. Ten ratios are sampled between $\rho_{\min}=\max(500\,\mathrm{kg}/m_{b,\max},0.05)$ and one, and Eq.~\eqref{eq21} is solved with $m_a=\rho m_{a,\max}$. Each transfer group also retains the boundary solution. Interior labels satisfy the SCP checks and the 500-kg terminal-mass acceptance threshold. Thus, these datasets cover the mass-dependent propellant response within each transfer condition, whereas the capacity datasets describe its boundary.

\begin{figure}[!h]
    \centering
    \begin{subfigure}[t]{0.45\linewidth}
        \centering
        \includegraphics[width=\linewidth]{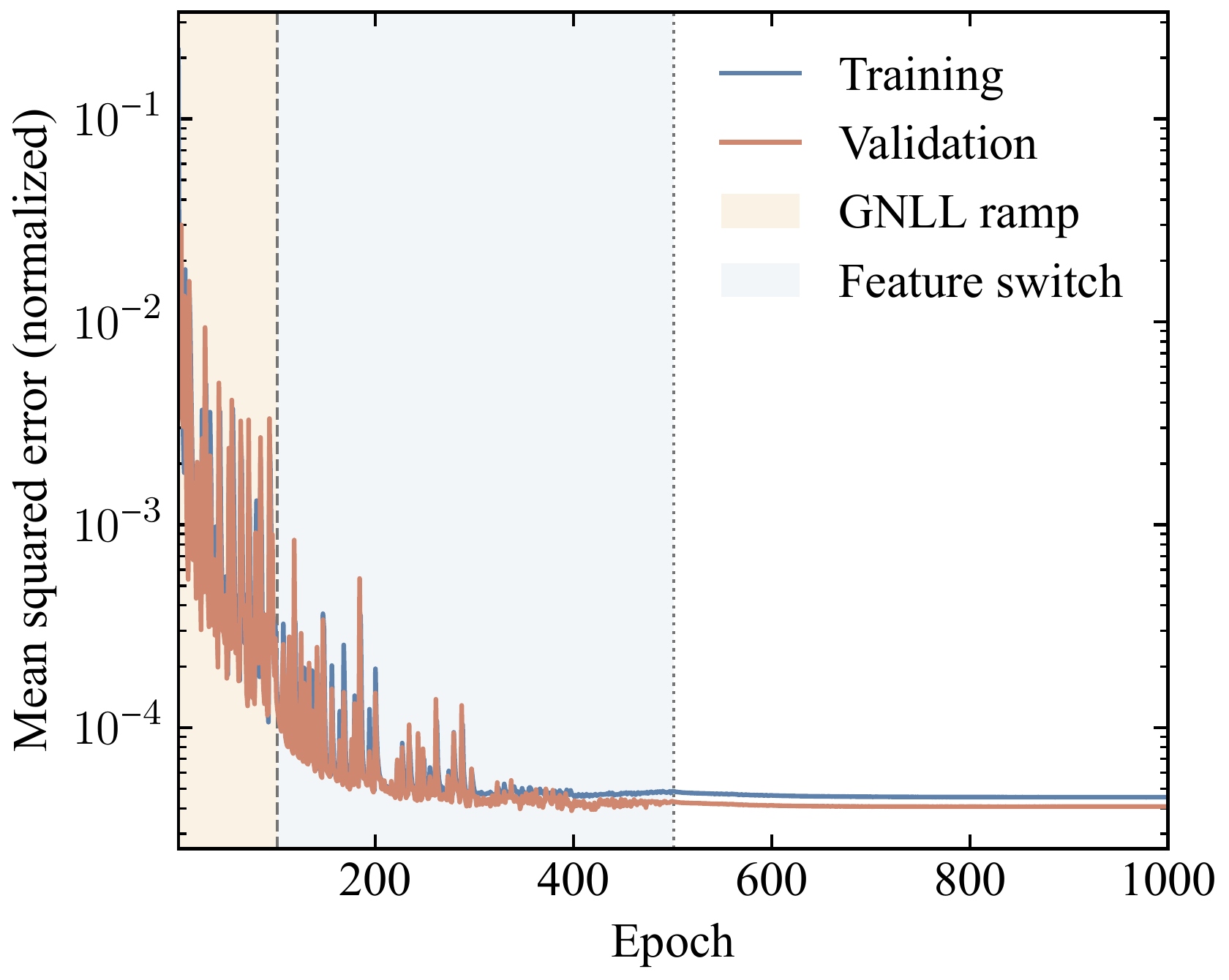}
        \caption{Training convergence.}
        \label{fig2.1}
    \end{subfigure}\hfill
    \begin{subfigure}[t]{0.45\linewidth}
        \centering
        \includegraphics[width=\linewidth]{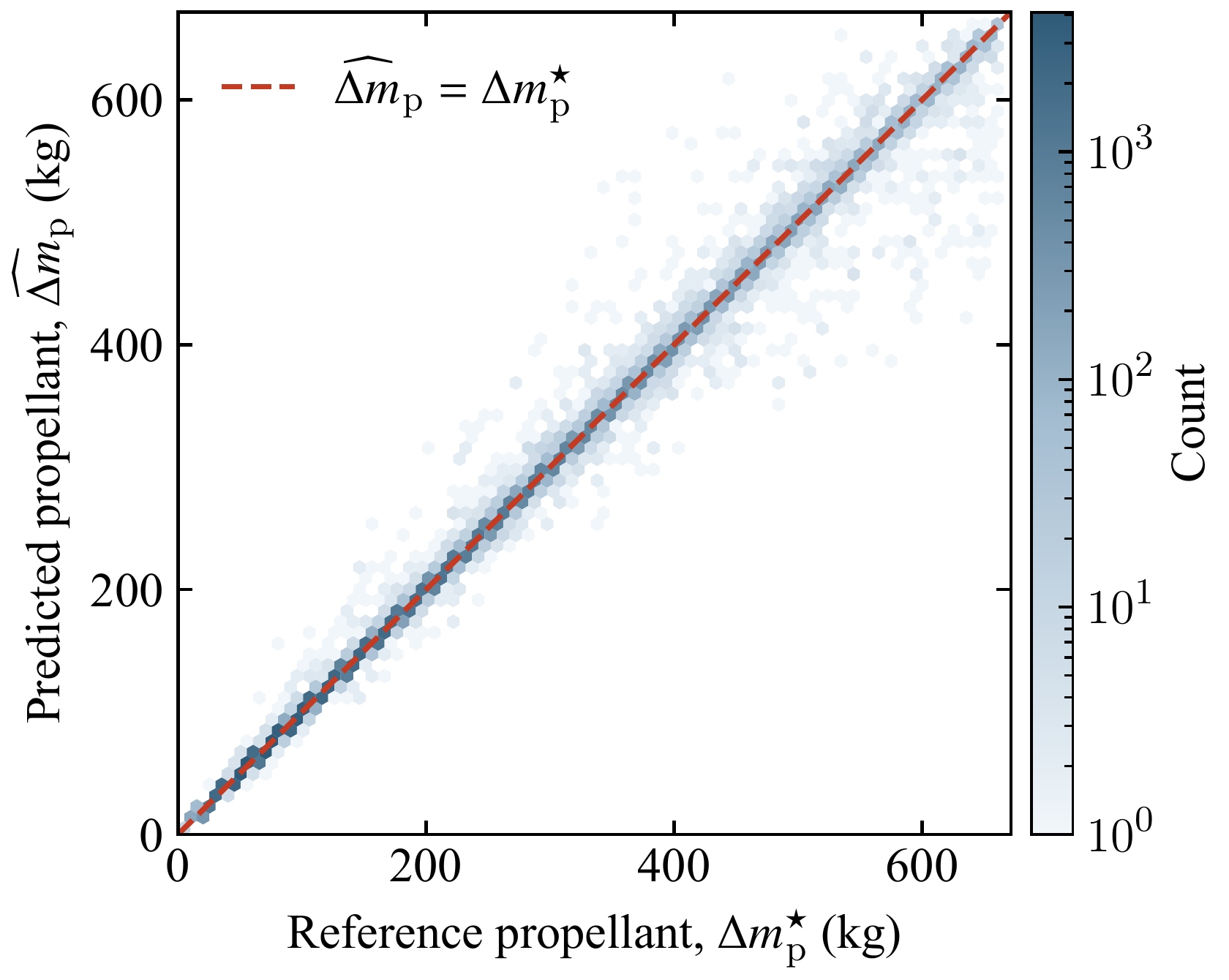}
        \caption{Prediction accuracy.}
        \label{fig2.2}
    \end{subfigure}
    \par\smallskip
    \begin{subfigure}[t]{0.45\linewidth}
        \centering
        \includegraphics[width=\linewidth]{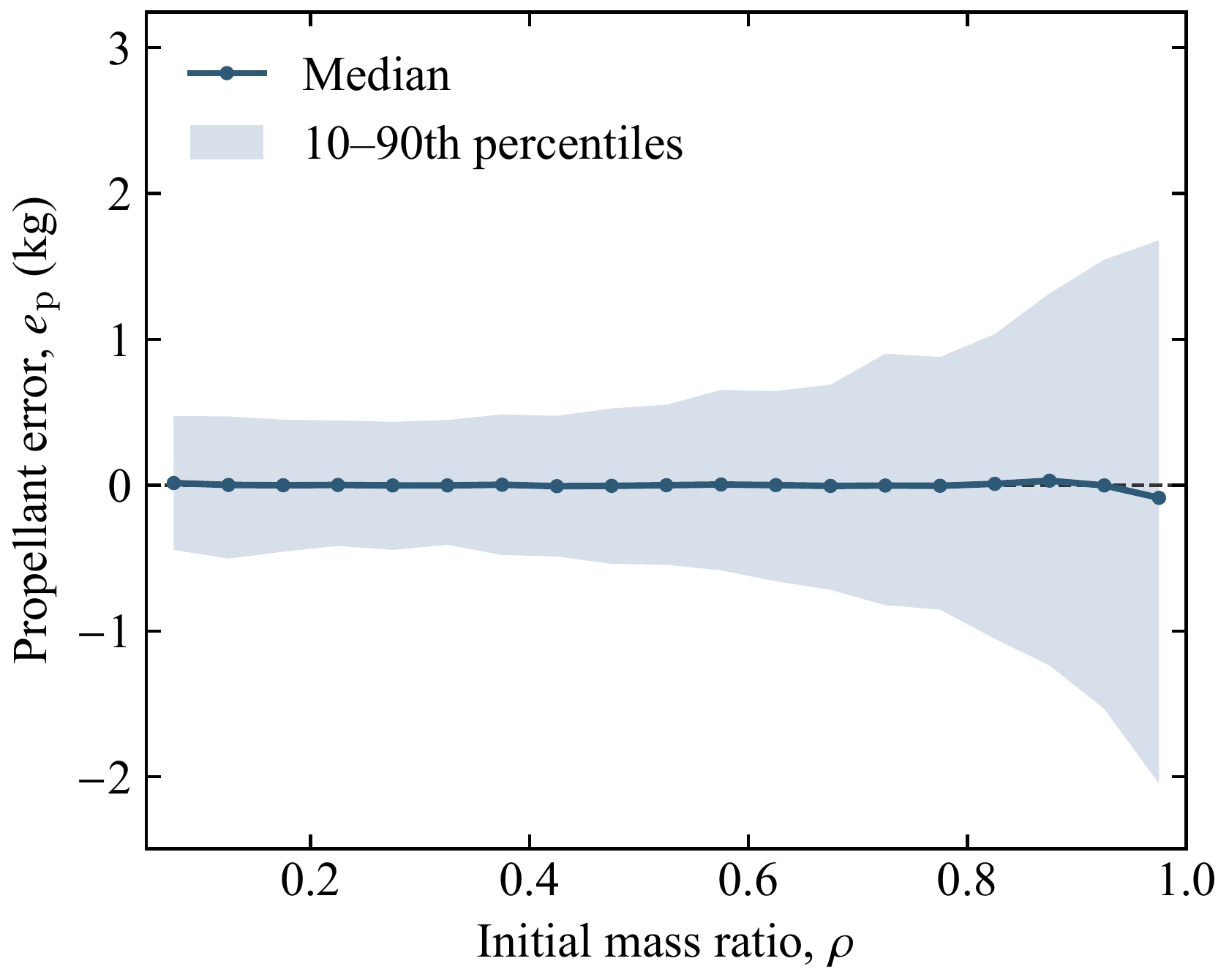}
        \caption{Mass-ratio dependence.}
        \label{fig2.3}
    \end{subfigure}\hfill
    \begin{subfigure}[t]{0.45\linewidth}
        \centering
        \includegraphics[width=\linewidth]{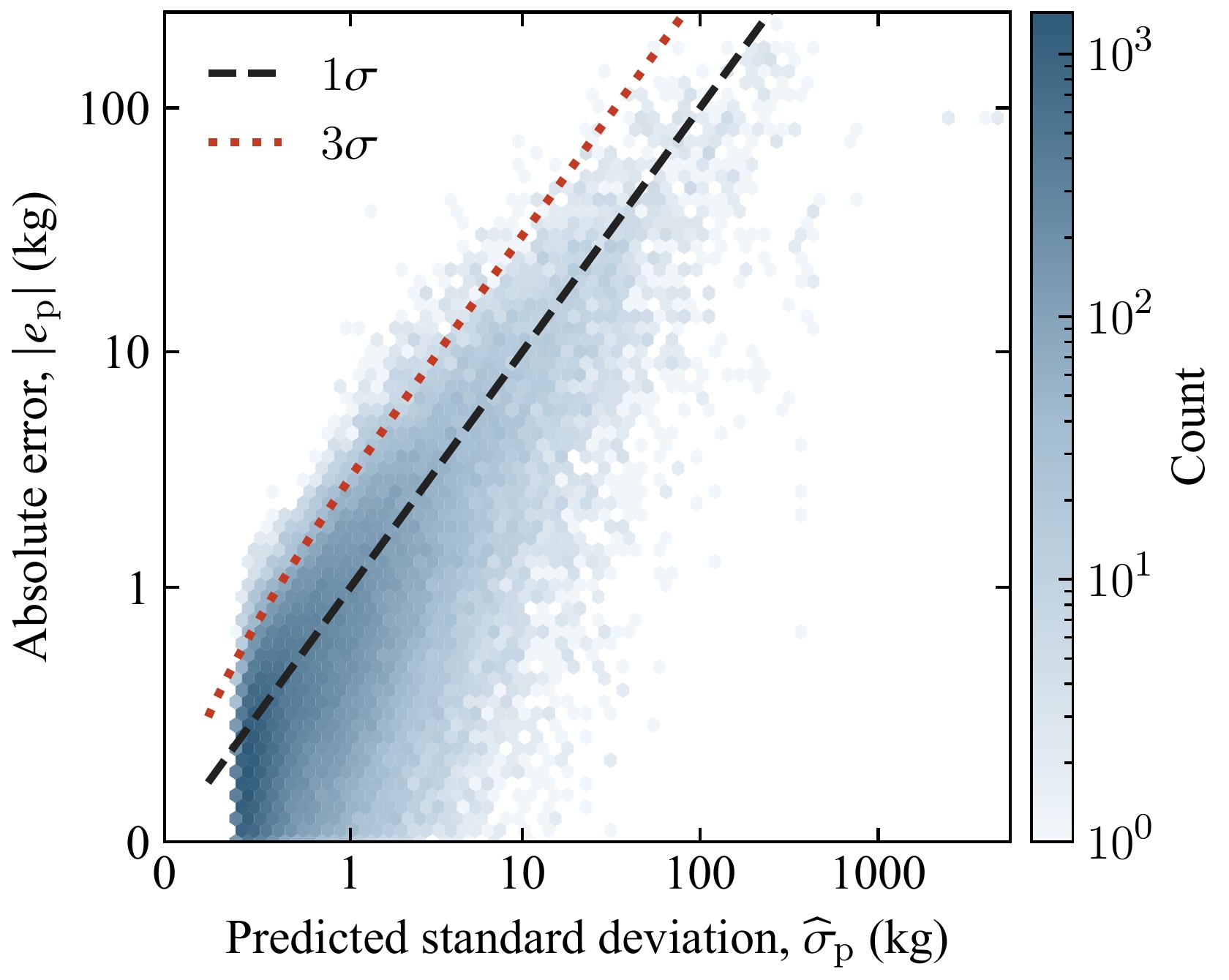}
    \caption{Surrogate-error diagnostic.}
        \label{fig2.4}
    \end{subfigure}
    \caption{Training and prediction performance of the propellant-consumption MANN.}
    \label{fig2}
\end{figure}

The propellant model uses the shared optimizer settings and Eq.~\eqref{eq14} with $E_{\mathrm{w}}=100$. To expose the model to the capacity estimates available during sequence search, training progressively replaces $(\rho,q)$ with $(\widehat{\rho},\widehat{q})$ using probability
\begin{equation}
    p_e=
    \begin{cases}
        0, & e<100,\\
        (e-100)/400, & 100\le e<500,\\
        1, & e\ge500 .
    \end{cases}
    \label{eq23}
\end{equation}
Checkpoint selection begins after both transitions are complete, $w_e=p_e=1$. The independent test results use predicted capacity features and therefore include their contribution to the propellant error. The training history of the selected propellant model is shown in Fig.~\ref{fig2}(\subref{fig2.1}).

\subsubsection{Prediction Performance}
\label{sec3.3.4}

Table~\ref{tab4} compares validation-selected DNN, KAN, and MANN configurations at 250,000 development records. The two MAE columns correspond to $(m_{a,\max},m_{b,\max})$ for initial-mass prediction and $(\Delta m_{\mathrm p},m_b)$ for propellant prediction.

\begin{table}[!h]
\caption{Single-leg prediction errors for validation-selected network configurations.}
\label{tab4}
\centering
\begin{tabular}{@{}lrrrr@{}}
\hline\hline
Model and configuration & Records & GNLL & MAE 1, kg & MAE 2, kg\\
\midrule
\multicolumn{5}{@{}l}{Maximum admissible initial mass}\\
DNN, PReLU, $6\times256$ & 250k & -4.9376 & 226.15 & 226.70\\
KAN, $4\times256$ & 250k & -4.5386 & 262.45 & 262.25\\
MANN, $6/64$ & 250k & -6.2730 & 62.88 & 63.28\\
MANN, $6/64$ & 1M & -7.0191 & 41.45 & 40.97\\
\midrule
\multicolumn{5}{@{}l}{Propellant consumption}\\
DNN, Swish, $4\times256$ & 250k & -5.7648 & 7.8902 & 8.0210\\
KAN, $4\times256$ & 250k & -6.1509 & 5.0629 & 5.5919\\
MANN, $10/64$ & 250k & -6.5948 & 1.9314 & 6.3467\\
MANN, $10/64$ & 1M & -7.3431 & 1.1713 & 3.3569\\
\hline\hline
\end{tabular}
\end{table}

For the validation-selected 250,000-record configurations, MANN has the lowest independent-test propellant MAE in Table~\ref{tab4}, 1.9314 kg, while KAN has the lowest independent-test terminal-mass MAE, 5.5919 kg. The million-record MANN reduces these errors to 1.1713 and 3.3569 kg, with corresponding mean relative errors of 0.5381 and 0.1564\%.

Predicted propellant consumption closely follows the optimization-derived labels across the 100,000 test samples [Fig.~\ref{fig2}(\subref{fig2.2})]. The signed error $e_{\mathrm{p}}=\widehat{\Delta m}_{\mathrm{p}}-\Delta m_{\mathrm{p}}^\star$ has a median near zero over $\rho\in[0.05,1]$, while its 10th--90th percentile band widens near the capacity boundary [Fig.~\ref{fig2}(\subref{fig2.3})].

Capacity-related features improve the propellant estimate. In the $6/64$ input comparison, adding the mass ratio and capacity-related mass-loss feature reduces MAE from 5.5356 to 2.8181~kg (Table~\ref{tabB1}). Among the tested configurations, the selected $10/64$ network has the lowest propellant MAE. Its propellant estimate is propagated during sequence search, whereas a different configuration has the lowest terminal-mass error (Table~\ref{tabB2}).

Table~\ref{tab5} summarizes the feature contributions under spherical coordinates, 250,000 development records, and the $6/64$ MANN sensitivity setting. The complete feature and coordinate scans remain in Appendix~\ref{sec7.1}--\ref{sec7.2}. Additional architecture, feature, training, and uncertainty analyses are provided in the Supplementary Material.

\begin{table}[!h]
\caption{Effect of selected physical and capacity-related features.}
\label{tab5}
\centering
\begin{tabular}{@{}llll@{}}
\hline\hline
Prediction task & Reduced input & Selected input & Target MAE, kg\\
\midrule
Maximum admissible initial mass & F0 & F11 & 993.36 $\rightarrow$ 62.88\\
Propellant consumption & F0 & F4 & 5.5356 $\rightarrow$ 2.8181\\
\hline\hline
\end{tabular}
\end{table}

For the selected model, standardized residuals are centered near zero, with spreads of 1.0569 and 1.0258 for propellant and terminal mass, respectively; their two-standard-deviation empirical coverages are 94.00 and 94.63\%. The predicted propellant standard deviation $\widehat{\sigma}_{\mathrm{p}}$ increases with $|e_{\mathrm{p}}|$ [Fig.~\ref{fig2}(\subref{fig2.4})], and empirical coverage is 70.83, 94.00, and 98.79\% at one, two, and three predicted standard deviations.
The maximum-admissible-initial-mass prediction supplies the capacity-related features required by the propellant model, whereas the propellant prediction propagates the branch mass used by the resource filter. Section~\ref{sec4} defines this integration, and Sec.~\ref{sec5} evaluates the resulting mission trajectories after low-thrust refinement.

\section{Methodology}
\label{sec4}

The solution method separates sequence search from continuous trajectory optimization. A Lambert-based ranking rule orders partial sequences, while the two MANN models estimate transfer capacity and propagate spacecraft mass. Resource filtering acts before survivor allocation, so extensions that fail the filter do not consume the limited beam budget. SCP then refines the retained sequences and checks their low-thrust trajectories. The following subsections define candidate expansion and allocation, the resource filter, trajectory refinement, and their combined procedure.

\subsection{Lambert-Based Beam Search}
\label{sec4.1}

\subsubsection{Candidate Expansion}
\label{sec4.1.1}

A node at encounter depth $d$ represents the partial sequence $\boldsymbol{s}_{0:d}$, its epochs $\boldsymbol{t}_{0:d}$, the current mission phase, and the accumulated cost and mass states. The sequence determines the visited-target set and the remaining admissible encounters. An extension appends a target--epoch pair $(s_{d+1},t_{d+1})$ satisfying the visit rules in Sec.~\ref{sec2.3}. Arrival epochs are drawn from a discrete grid subject to transfer-duration bounds and the mission window. For a prescribed visit count, a branch is also required to leave sufficient time for the remaining encounters.

The transfer from event $k$ to event $k+1$ is initially evaluated by solving a Lambert problem between the two target positions. With Lambert departure and arrival velocities denoted by $\boldsymbol{v}_{k,\mathrm{L}}^{\mathrm{dep}}$ and $\boldsymbol{v}_{k+1,\mathrm{L}}^{\mathrm{arr}}$, the impulsive rendezvous cost is
\begin{equation}
    \Delta v_k^{\mathrm{L}}
    =
    \left\|\boldsymbol{v}_{k,\mathrm{L}}^{\mathrm{dep}}
    -\boldsymbol{v}_{s_k}(t_k)\right\|
    +
    \left\|\boldsymbol{v}_{s_{k+1}}(t_{k+1})
    -\boldsymbol{v}_{k+1,\mathrm{L}}^{\mathrm{arr}}\right\|.
    \label{eq24}
\end{equation}
For the rendezvous missions, the corresponding recursive mass estimate is
\begin{equation}
    m_{k+1}^{\mathrm{L},-}
    =
    m_k^{\mathrm{L},+}
    \exp\!\left(-\frac{\Delta v_k^{\mathrm{L}}}{g_0I_{\mathrm{sp}}}\right),
    \qquad
    m_{k+1}^{\mathrm{L},+}
    =
    m_{k+1}^{\mathrm{L},-}+\Delta m_{k+1}^{\mathrm{event}}.
    \label{eq25}
\end{equation}
Candidates with larger $m^{\mathrm{L}}$ are preferred. For a common initial mass and no event-related mass changes, this ordering is equivalent to minimizing the accumulated cost $\sum_k\Delta v_k^{\mathrm{L}}$. The deployment--collection search uses the accumulated Lambert cost directly, with payload changes treated in the resource calculation. These case-specific metrics rank candidates during search. Maximum-visit trajectories are compared first by visit count and then by terminal mass; deployment--collection trajectories are compared by returned mined mass.

Candidate generation first enforces visit and time restrictions, Lambert solvability, and baseline resource bounds. A preliminary Lambert-based truncation limits the candidate pool passed to resource evaluation. Resource filtering and local pruning then precede survivor allocation. The event grid and preliminary candidate limit are specified in the Supplementary Material.

\subsubsection{Survivor Allocation}
\label{sec4.1.2}

Let $\mathcal{B}_d$ denote the retained population at depth $d$, and let $\mathcal{C}_{d+1}$ contain its admissible extensions. The depth-synchronous search can be expressed as
\begin{equation}
\begin{aligned}
    \mathcal{D}_{d+1}&=\mathcal{F}(\mathcal{C}_{d+1}),\\
    \mathcal{B}_{d+1}
    &=\mathcal{S}_{B}^{\alpha}
    \!\left[\mathcal{P}(\mathcal{D}_{d+1});\,\mathcal{D}_{d+1}\right],
    \qquad |\mathcal{B}_{d+1}|\leq B.
\end{aligned}
    \label{eq26}
\end{equation}
Here, $\mathcal{F}$ is the resource filter defined in Sec.~\ref{sec4.2}, or the identity operation for the Lambert baseline; $\mathcal{D}_{d+1}$ is the eligible pool; $\mathcal{P}$ denotes local pruning; and $\mathcal{S}_{B}^{\alpha}$ allocates at most $B$ survivors according to architecture $\alpha$. Filtering tests predicted low-thrust resources, whereas local pruning limits redundant states associated with a common terminal event or route. The second argument of $\mathcal{S}_{B}^{\alpha}$ supplies the original eligible pool for backfilling where enabled: unused slots are filled from its remaining candidates in Lambert order.

The first three architectures share depth-synchronous expansion and differ in the distribution of survivor slots. Their diversity groups are defined by the ordered body sequence $\boldsymbol{s}_{0:d}$, excluding encounter epochs. Consequently, different timing realizations of the same route compete within one group, whereas different routes belong to different groups.

\textit{Standard beam search.} After local pruning, the candidates are globally ordered by the Lambert metric and the first $B$ are retained. This policy concentrates the available slots on the highest-ranked candidates without an additional route-diversity restriction.

\textit{Sequence-Capped beam search.} Candidates are scanned in Lambert order with an initial cap $c_0$ on the number of survivors per body-sequence group. The positive integer $c_0$ controls the initial concentration of slots within each group; its value is specified in Sec.~\ref{sec5}. If slots remain vacant, the cap is increased by one in successive passes until the beam is filled or the eligible pool is exhausted. The policy limits early concentration on a single route while allowing unused capacity to be reassigned.

\textit{Round-Robin beam search.} Candidates are ranked within each body-sequence group, and slots are allocated in successive rounds, with at most one candidate drawn from each nonempty group per round. A deterministic Lambert-based ordering resolves competition among groups. Unlike a fixed cap, round-robin allocation first distributes slots across the available routes before admitting additional timing realizations.

\textit{Time-Window beam search.} This architecture changes the expansion schedule as well as the selection pool. Eligible nodes are stored in a queue $\mathcal{Q}$ according to their elapsed mission time $\tau$. At each iteration, the earliest window is extracted as
\begin{equation}
    \mathcal{Q}_{\varepsilon}
    =
    \{b\in\mathcal{Q}:\tau_{\min}\leq\tau(b)<\tau_{\min}+\varepsilon\},
    \qquad
    \tau_{\min}=\min_{b\in\mathcal{Q}}\tau(b).
    \label{eq27}
\end{equation}
The window width is $\varepsilon$. Local pruning by terminal event and the Lambert-based ranking rule select at most $B$ nodes from this window for expansion. The remaining window nodes are discarded, while nodes outside the window remain queued. Eligible nonterminal children are inserted into the queue, and completed sequence candidates are archived. Since one window may contain several encounter depths, $B$ bounds the number expanded per window rather than the total queue size.

These policies separate two choices: how slots are distributed among competing routes and which nodes compete at the same selection step. Table~\ref{tab6} summarizes this distinction after the common candidate-generation operations.
\begin{table}[!h]
\caption{Beam survivor allocation and expansion scheduling.}
\label{tab6}
\centering
\begin{tabular}{lll}
\hline\hline
Architecture & Allocation rule & Competing population \\
\hline
Standard & Global Lambert truncation & Candidates at a common depth \\
Sequence-Capped & Initial per-sequence cap $c_0$, with relaxation & Candidates at a common depth \\
Round-Robin & Successive rounds across route groups & Candidates at a common depth \\
Time-Window & Local pruning and Lambert truncation & Nodes in the earliest time window \\
\hline\hline
\end{tabular}
\end{table}

\subsection{MANN-Based Resource Filtering}
\label{sec4.2}

An impulsive ranking does not account explicitly for the thrust-limited transfer capacity. The resource filter introduces this information through the MIM and propellant models in Sec.~\ref{sec3}. For each proposed asteroid-to-asteroid extension, the endpoint states are rotated into the canonical departure frame and the transfer features are constructed. The capacity estimate is the fused MIM mean $\widehat{m}_{a,\max,k}^{\mathrm{fus}}$ from Eq.~\eqref{eq20}; propellant consumption is evaluated from the direct propellant output at the current branch mass.

A separate state $\widehat{m}^{\mathrm{M}}$ carries the MANN-based mass estimate along the branch. For forward propagation, the capacity margin is
\begin{equation}
    g_k^{\mathrm{cap}}
    =
    \widehat{m}_{a,\max,k}^{\mathrm{fus}}
    -
    \widehat{m}_k^{\mathrm{M},+},
    \qquad
    g_k^{\mathrm{cap}}\geq 0.
    \label{eq28}
\end{equation}
The two terms in Eq.~\eqref{eq28} have distinct dependencies: the fused MIM estimate is determined by the current transfer conditions, whereas the propagated spacecraft mass reflects the preceding transfers and event-related mass changes. Their difference converts a single-leg capacity prediction into a sequence-dependent eligibility test, so candidates arriving at the same transfer conditions with different mass histories are evaluated against the same predicted capacity boundary.
For forward sequence construction with an available physical branch mass, a negative predicted capacity margin removes the extension from the population eligible for beam allocation. A negative capacity margin is interpreted as predicted low-thrust infeasibility for search purposes; it is a surrogate-based screening decision rather than a mathematical certificate of infeasibility. For an extension that passes, the propellant model uses $\widehat{m}_k^{\mathrm{M},+}$ and the predicted capacity-related features in Sec.~\ref{sec3.3} to obtain $\widehat{\Delta m}_{\mathrm{p},k}$. The arrival and post-event masses are then
\begin{equation}
    \widehat{m}_{k+1}^{\mathrm{M},-}
    =
    \widehat{m}_k^{\mathrm{M},+}
    -
    \widehat{\Delta m}_{\mathrm{p},k},
    \qquad
    \widehat{m}_{k+1}^{\mathrm{M},+}
    =
    \widehat{m}_{k+1}^{\mathrm{M},-}
    +
    \Delta m_{k+1}^{\mathrm{event}}.
    \label{eq29}
\end{equation}
The deployment--collection case uses a task-specific reverse collection construction with a representative-mass proxy, described in Sec.~\ref{sec5.4}. After the deployment and collection branches are joined, the complete resource history is reevaluated forward in physical time before Earth-return closure and SCP refinement.

For rendezvous missions, the arrival estimate must satisfy $\widehat{m}_{k+1}^{\mathrm{M},-}\geq m_{\mathrm{dry}}$. In deployment--collection missions, these updates define the forward deployment propagation and the post-join forward replay, including miner release and material collection. The full payload-dependent constraints of Sec.~\ref{sec2.3} are enforced in trajectory refinement.

The distinction between $m^{\mathrm{L}}$ and $\widehat{m}^{\mathrm{M}}$ preserves the role of each model. The former supplies the rendezvous ranking metric; the latter supplies the mass entering subsequent low-thrust predictions and eligibility tests. In Eq.~\eqref{eq26}, filtering therefore changes the argument of the survivor-allocation operator rather than its ranking rule. For Time-Window beam search, the same resource test is applied to children before queue insertion, so they have already passed filtering when their time window is selected.

The resource filter uses predicted means for screening and recursive mass propagation, whereas the predicted standard deviations are retained as surrogate-error diagnostics. Filter decisions determine beam eligibility, and final low-thrust trajectory acceptance is determined by SCP. Equations~\eqref{eq28} and \eqref{eq29} define forward propagation; Sec.~\ref{sec5.4} introduces the reverse-construction approximation used for collection.

\subsection{Trajectory Refinement}
\label{sec4.3}

Each retained sequence defines a continuous low-thrust optimal-control problem, initialized at the search-generated encounter epochs. SCP solves a sequence of convex approximations to its nonlinear dynamics and constraints, following the successive-convexification approach \cite{Mao2016SCvx}. The state contains nondimensional position, velocity, and logarithmic mass, $\zeta=\log(m/m_\ast)$. Piecewise-constant control variables describe acceleration and its magnitude. Lambert-based trajectories initialize the state history, and the controls obtained in an initial pass initialize the subsequent refinement. The objective is the corresponding mission objective in Sec.~\ref{sec2}, not the Lambert search cost.

Let $\boldsymbol{y}_{k,j}$ be the scaled state at node $j$ of leg $k$, $\boldsymbol{u}_{k,j}$ its control variables, and $h_{k,j}$ the integration-interval duration. The nonlinear interval flow is $\boldsymbol{\Phi}_{k,j}$. At iteration $\ell$, linearization about the reference trajectory gives
\begin{equation}
    \boldsymbol{y}_{k,j+1}
    ={}
    \overline{\boldsymbol{\Phi}}_{k,j}
    +A_{k,j}\bigl(\boldsymbol{y}_{k,j}-\overline{\boldsymbol{y}}_{k,j}\bigr)
    +B_{k,j}\bigl(\boldsymbol{u}_{k,j}-\overline{\boldsymbol{u}}_{k,j}\bigr)+D_{k,j}\bigl(h_{k,j}-\overline{h}_{k,j}\bigr)
    +\boldsymbol{\nu}_{k,j}.
\label{eq30}
\end{equation}
Overbars denote the reference trajectory, and $A_{k,j}$, $B_{k,j}$, and $D_{k,j}$ are flow sensitivities with respect to state, control, and duration. In the rendezvous formulation, penalized virtual controls $\boldsymbol{\nu}_{k,j}$ relax the affine dynamics. The deployment--collection formulation imposes these dynamics directly, with $\boldsymbol{\nu}_{k,j}=0$. Encounter ephemerides and logarithmic event-mass links are linearized with respect to their decision variables. Trust regions bound changes in state and time, while convex state constraints restrict the admissible iterates \cite{Mao2017StateConstraints}.

Writing $\boldsymbol{z}$ for the stacked trajectory decision variables and $\boldsymbol{\xi}$ for the active relaxation variables, a convex subproblem has the form
\begin{equation}
\begin{aligned}
    \underset{\boldsymbol{z},\,\boldsymbol{\xi}}{\operatorname{maximize}}
    \quad&
    \widetilde{J}_{\ell}(\boldsymbol{z})
    -\boldsymbol{\lambda}^{\mathsf T}|\boldsymbol{\xi}|\\
    \text{subject to}\quad&
    \text{affine propagation in Eq.~\eqref{eq30}},\\
    &\text{linearized encounter and event-mass relations},\\
    &\boldsymbol{z}\in\mathcal{C}_{\ell}\cap\mathcal{T}_{\ell}.
\end{aligned}
\label{eq31}
\end{equation}
Here, $\mathcal{C}_{\ell}$ contains convex thrust, mass, and time constraints, and $\mathcal{T}_{\ell}$ is the trust region. The vector $\boldsymbol{\xi}$ collects the active relaxation variables with positive penalty weights $\boldsymbol{\lambda}$. The acceleration-norm bound is a second-order cone, and the mass-dependent thrust bound is approximated about the reference logarithmic mass \cite{Wang2018Convex}. For rendezvous missions, $\widetilde J_\ell$ maximizes terminal logarithmic mass, which preserves the terminal-mass ordering. Deployment--collection refinement maximizes returned mined mass. The visit count is fixed during each continuous solve in the maximum-visit case.

The optimized controls and durations are propagated through the nonlinear equations after each convex solve. Endpoint discrepancies measure consistency between the resulting trajectory and the imposed encounters; logarithmic-mass discrepancies are also checked in deployment--collection refinement. The propagated solution defines the next reference. Iteration stops at the prescribed tolerances, iteration limit, or solver termination. The acceptance criteria are defined in Sec.~\ref{sec5.1}.

Fixed-time (FT) refinement imposes $\boldsymbol{t}=\boldsymbol{t}^{\mathrm B}$, where $\boldsymbol{t}^{\mathrm B}$ denotes the search-generated epochs. Variable-time (VT) refinement retains the sequence but releases the encounter epochs within the mission bounds. Time-adjusted ephemerides and interval durations enter the same linearization. Section~\ref{sec5} specifies the numerical passes used for each case and reports each refinement mode separately.

\subsection{Overall Procedure}
\label{sec4.4}

Algorithm~\ref{alg2} summarizes the search and refinement sequence. The terminal archive $\mathcal{H}$ contains candidate sequences and epochs, and $\mathcal{R}$ contains their refinement results.
\begin{algorithm}[!h]
\caption{Lambert-ranked beam search with MANN resource filtering and SCP refinement}
\label{alg2}
\begin{algorithmic}[1]
\Require Ephemerides, mission constraints, event grid, architecture $\alpha$, width $B$, window width $\varepsilon$, trained models, and refinement settings.
\Ensure Candidate archive $\mathcal{H}$ and refinement results $\mathcal{R}$.
\State Initialize departure states; $\mathcal{H}\gets\varnothing$, $\mathcal{R}\gets\varnothing$.
\State Store the initial population in $\mathcal{B}$, or in $\mathcal{Q}$ for Time-Window.
\While{an expandable population remains}
    \If{$\alpha=\text{Time-Window}$}
        \State Extract $\mathcal{Q}_{\varepsilon}$; $\mathcal{Q}\gets\mathcal{Q}\setminus\mathcal{Q}_{\varepsilon}$. \Comment{Eq.~\eqref{eq27}}
        \State $\mathcal{E}\gets\mathcal{S}^{\mathrm{Standard}}_B[\mathcal{P}(\mathcal{Q}_{\varepsilon})]$.
    \Else
        \State $\mathcal{E}\gets\mathcal{B}$.
    \EndIf
    \State Generate admissible extensions $\mathcal{C}$ from $\mathcal{E}$.
    \State Evaluate Lambert costs and resource bounds; retain the preliminary Lambert-ranked pool.
    \State $\mathcal{C}\gets\mathcal{F}(\mathcal{C})$; use the identity for the baseline. \Comment{Sec.~\ref{sec4.2}}
    \If{$\alpha=\text{Time-Window}$}
        \State Archive completed candidates; queue eligible nonterminal children.
    \Else
        \State $\mathcal{B}\gets\mathcal{S}^{\alpha}_B[\mathcal{P}(\mathcal{C});\,\mathcal{C}]$. \Comment{Eq.~\eqref{eq26}}
        \State Update $\mathcal{H}$ with completed candidates; retain expandable states in $\mathcal{B}$.
    \EndIf
\EndWhile
\State Apply required forward resource evaluation and Earth-return closure to $\mathcal{H}$.
\State Select the prescribed number of candidates from $\mathcal{H}$ in benchmark order.
\For{each selected candidate}
    \State Solve FT and VT; store each trajectory and status. \Comment{Sec.~\ref{sec4.3}}
    \State Check acceptance and append stage-specific objective values to $\mathcal{R}$.
\EndFor
\State \Return $\mathcal{H},\mathcal{R}$.
\end{algorithmic}
\end{algorithm}

Earth-boundary transfers are solved separately from the asteroid-to-asteroid surrogate to provide departure seeds or close return legs before full-mission refinement. Mission-specific phase transitions and boundary conditions are described in Sec.~\ref{sec5}.

\section{Numerical Results}
\label{sec5}

The numerical experiments examine how MANN resource filtering changes the sequences retained by Lambert-ranked beam search and their quality after low-thrust refinement. Three cases instantiate the mission classes in Sec.~\ref{sec2}: fixed-visit rendezvous, maximum-visit rendezvous, and deployment--collection.

\subsection{Benchmark Setup}
\label{sec5.1}

All three cases use the same set $\mathcal{A}$ of 15 candidate asteroids from the GTOC12 catalogue \cite{Zhang2025GTOC12Design}, selected using the clustering-based subset strategy developed by the team “TheAntipodes” \cite{TheAntipodes2025}. Appendix~\ref{sec7.3} specifies their identifiers and catalogue orbital elements. The two rendezvous cases adopt GTOC4-inspired mission objectives but use these GTOC12-derived targets rather than the original GTOC4 catalogue. Target selection, encounter epochs, and mission-specific visit rules differ among the cases. The common target set tests different mission objectives and search policies; applicability to other targets is governed by the transfer sampling domain in Table~\ref{tab3}.

The spacecraft follows the heliocentric low-thrust model in Sec.~\ref{sec2.2}. Table~\ref{tab7} summarizes the shared vehicle parameters and the principal case-dependent settings. The rendezvous cases prescribe a launch mass of 3000~kg, whereas the deployment--collection case permits the launch mass to vary up to this value during full-mission refinement. The latter also includes miner release, material collection, and Earth return, as detailed in Sec.~\ref{sec5.4}.
\begin{table}[!h]
\caption{Shared parameters and case-dependent settings.}
\label{tab7}
\centering
\begin{tabular}{lccc}
\hline\hline
Parameter & Fixed-visit & Maximum-visit & Deployment--collection \\
\hline
Candidate asteroids & \multicolumn{3}{c}{15 shared targets} \\
Maximum thrust, N & \multicolumn{3}{c}{0.6} \\
Specific impulse, s & \multicolumn{3}{c}{4000} \\
Dry mass, kg & \multicolumn{3}{c}{500} \\
Launch mass, kg & 3000 & 3000 & $\leq 3000$ \\
Selected asteroids & 12 & Variable & 9 \\
Beam width & 30 & 50 & 50, 100 \\
Refinement candidates & \multicolumn{3}{c}{Up to 10 per comparison group} \\
\hline\hline
\end{tabular}
\end{table}

Within each case and architecture, the Lambert baseline and resource-filtered search use identical mission constraints, ranking rules, and survivor budgets. For the rendezvous cases, their only methodological difference is the mean-based resource filter in Sec.~\ref{sec4.2}. The deployment--collection case combines forward resource filtering with the proxy-based reverse collection construction described in Sec.~\ref{sec5.4}. Sequence-Capped search uses $c_0=2$. All four architectures are evaluated in the rendezvous cases, whereas the deployment--collection case uses the three depth-synchronous architectures.

Each comparison group refines up to ten highest-ranked candidate sequences. ``Valid'' counts candidate sequences whose SCP trajectories satisfy the acceptance criteria below; objective statistics include only these entries. In the maximum-visit case, the maximum SCP-validated visit count $N_{\max}$ is determined before terminal masses are compared.

All mission refinements use a logarithmic-mass state representation and second-order-cone subproblems. The characteristic scales are $r_\ast=1~\mathrm{AU}$, $v_\ast=\sqrt{\mu_\odot/r_\ast}$, $t_\ast=r_\ast/v_\ast$, and $m_\ast=3000~\mathrm{kg}$. For propagated endpoint residuals $\delta\boldsymbol r_k$ and $\delta\boldsymbol v_k$, define
\begin{equation}
 \epsilon_{\mathrm{rv}}
 =\max_k\left\{
 \frac{\|\delta\boldsymbol r_k\|_\infty}{r_\ast},
 \frac{\|\delta\boldsymbol v_k\|_\infty}{v_\ast}\right\}.
 \label{eq32}
\end{equation}
Earth-boundary velocity residuals exclude the permitted excess velocity. Rendezvous acceptance requires $\epsilon_{\mathrm{rv}}\le10^{-6}$ together with the dimensional encounter, thrust, mass, and time checks. Deployment--collection acceptance requires $\max(\epsilon_{\mathrm{rv}},\max_k|\delta\zeta_k|)\le10^{-6}$ and an admissible solver status, where $\delta\zeta_k$ compares the optimized terminal logarithmic mass with nonlinear propagation on leg $k$.

Fixed-time (FT) refinement preserves the search-generated epochs, whereas variable-time (VT) refinement also optimizes encounter epochs. In the deployment--collection case, an additional fixed-time refinement at VT-adjusted epochs is denoted VT$\rightarrow$FT.

\subsection{Fixed-Visit Case}
\label{sec5.2}

The fixed-visit case requires 12 distinct asteroid visits within eight years. The retained beam width is $B=30$, and Time-Window search uses $\varepsilon=50$ days.

Table~\ref{tab8} compares FT and VT refinement of the ten highest-ranked candidate sequences from each search; accepted-entry counts are therefore out of ten. Mean FT terminal mass is computed over accepted entries. The Spearman coefficient $\rho_s$ compares beam rank with descending FT mass rank, so positive values indicate agreement between the two orderings.
\begin{table}[!h]
\caption{Fixed-time and variable-time SCP results for the fixed-visit case.}
\label{tab8}
\centering
\begin{tabular}{llrrrrrr}
\hline\hline
 & & \multicolumn{4}{c}{FT} & \multicolumn{2}{c}{VT} \\
\cline{3-6}\cline{7-8}
Architecture & Method & Valid & Best, kg & Mean, kg & $\rho_s$ & Valid & Best, kg \\
\hline
Standard & Lambert & 8 & 1427.518 & 1236.651 & $-0.119$ & 8 & 1579.559 \\
Standard & MANN & 10 & 1433.115 & 1378.087 & $0.273$ & 10 & 1543.352 \\
Sequence-Capped & Lambert & 9 & 1427.518 & 1253.494 & $-0.267$ & 9 & 1579.559 \\
Sequence-Capped & MANN & 10 & 1433.115 & 1378.087 & $0.273$ & 10 & 1543.352 \\
Round-Robin & Lambert & 10 & 1433.115 & 1276.001 & $-0.067$ & 10 & 1579.559 \\
Round-Robin & MANN & 10 & 1460.833 & 1381.095 & $0.200$ & 10 & 1543.352 \\
Time-Window & Lambert & 10 & 1427.518 & 1227.346 & $0.212$ & 10 & 1518.087 \\
Time-Window & MANN & 10 & 1427.518 & 1355.609 & $0.867$ & 10 & 1512.814 \\
\hline\hline
\end{tabular}
\end{table}

At the search-generated epochs, the dominant effect of resource filtering differs by survivor-allocation architecture. Standard and Sequence-Capped search each yield ten accepted FT trajectories after filtering. In Round-Robin search, the best terminal mass increases from 1433.115 to 1460.833~kg. For Time-Window search, the best mass is unchanged, but the mean terminal mass and rank correlation rise from 1227.346 to 1355.609~kg and from 0.212 to 0.867, respectively.

\begin{figure}[!h]
\centering
\includegraphics[width=0.85\linewidth]{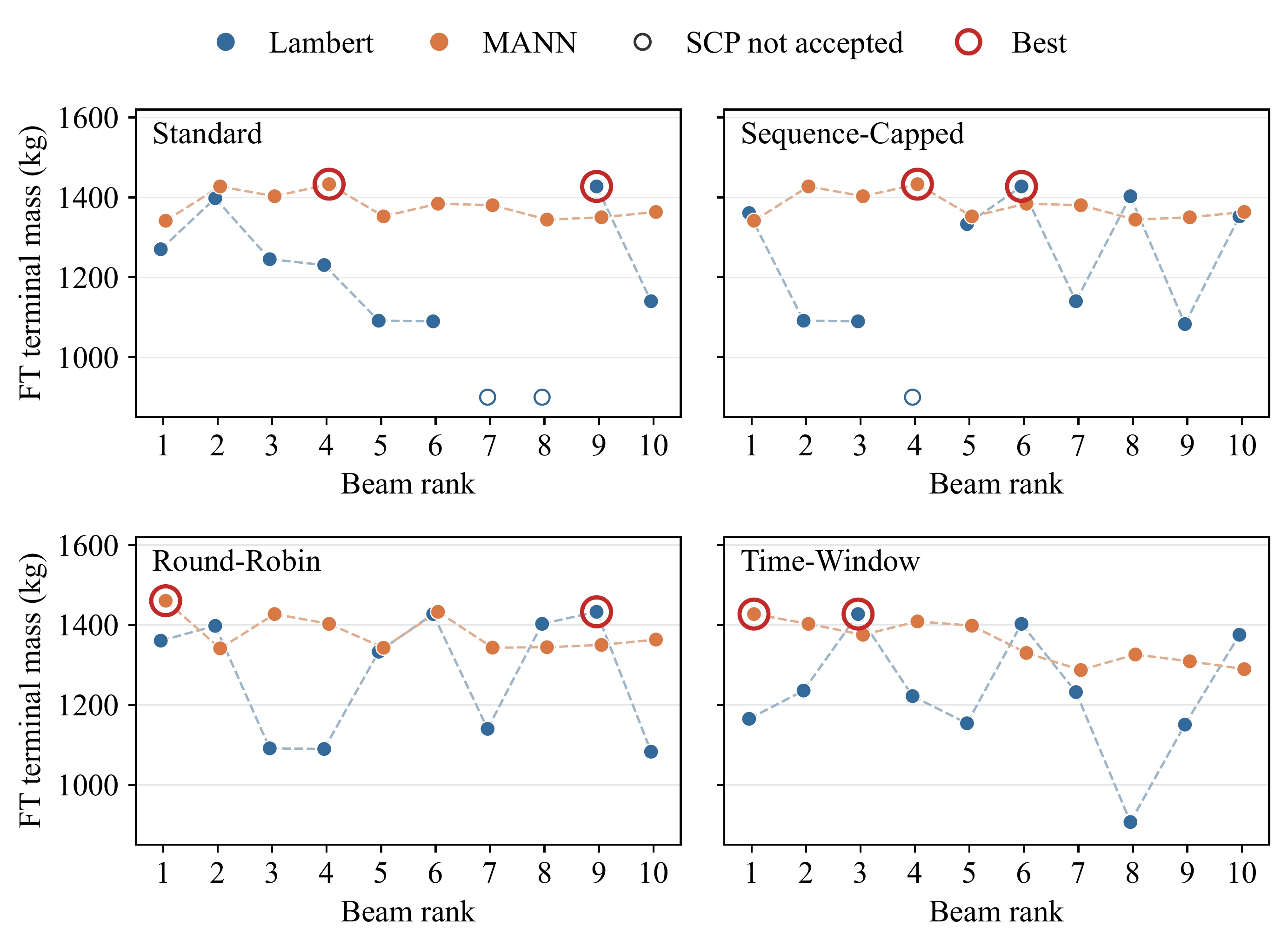}
\caption{FT terminal mass by beam rank. Rings mark best accepted entries; open symbols at 900~kg denote entries without accepted SCP trajectories.}
\label{fig3}
\end{figure}

Filtering also moves high-quality accepted FT trajectories to higher search ranks [Fig.~\ref{fig3}]. In Round-Robin and Time-Window search, the best accepted FT trajectory becomes the first-ranked candidate after filtering. For Time-Window search, this ordering improvement occurs without a change in the best terminal mass.

\begin{figure}[!h]
\centering
\begin{subfigure}[t]{0.49\linewidth}
\centering
\includegraphics[width=\linewidth]{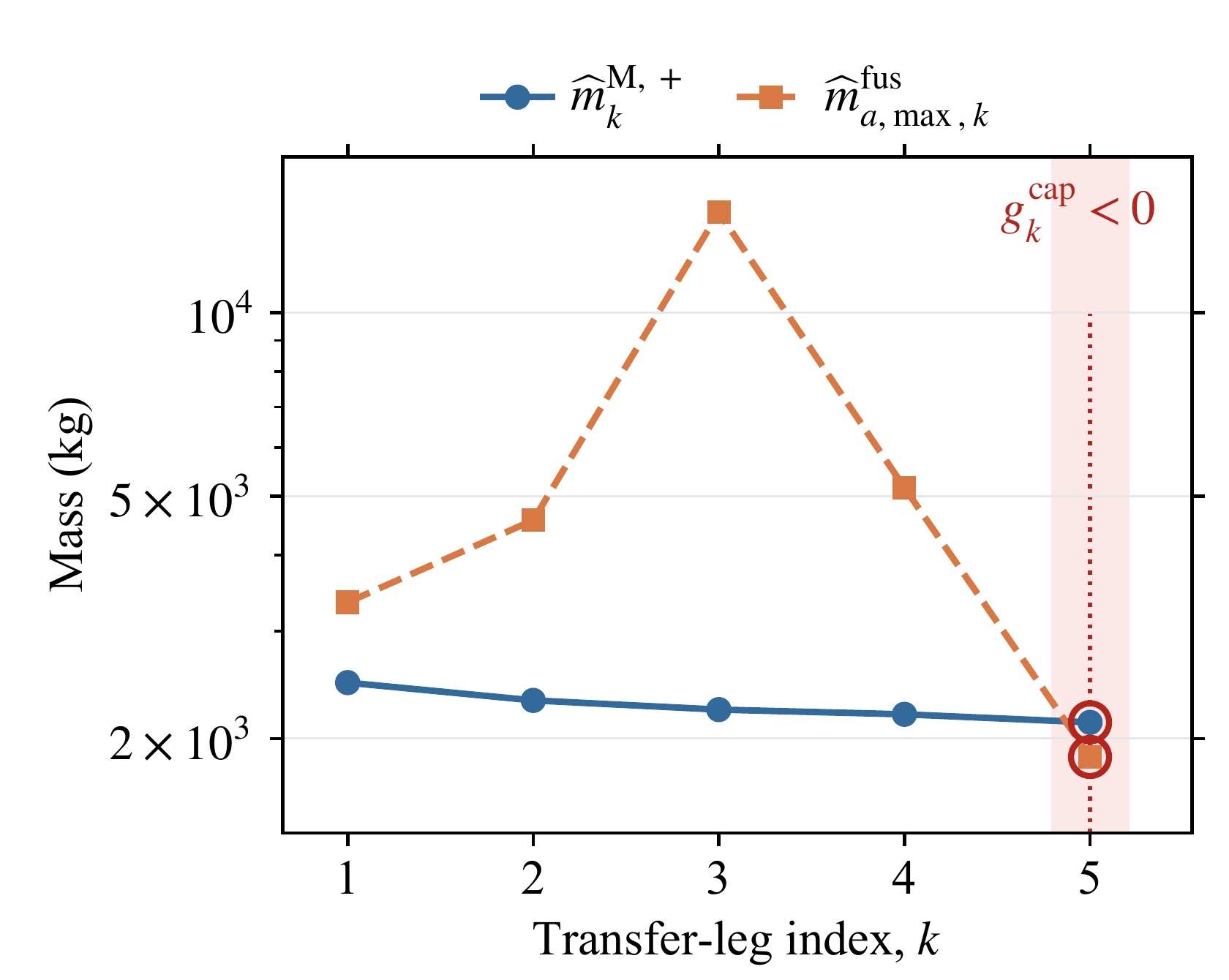}
\caption{Standard, rank 8}
\label{fig4.1}
\end{subfigure}\hfill
\begin{subfigure}[t]{0.49\linewidth}
\centering
\includegraphics[width=\linewidth]{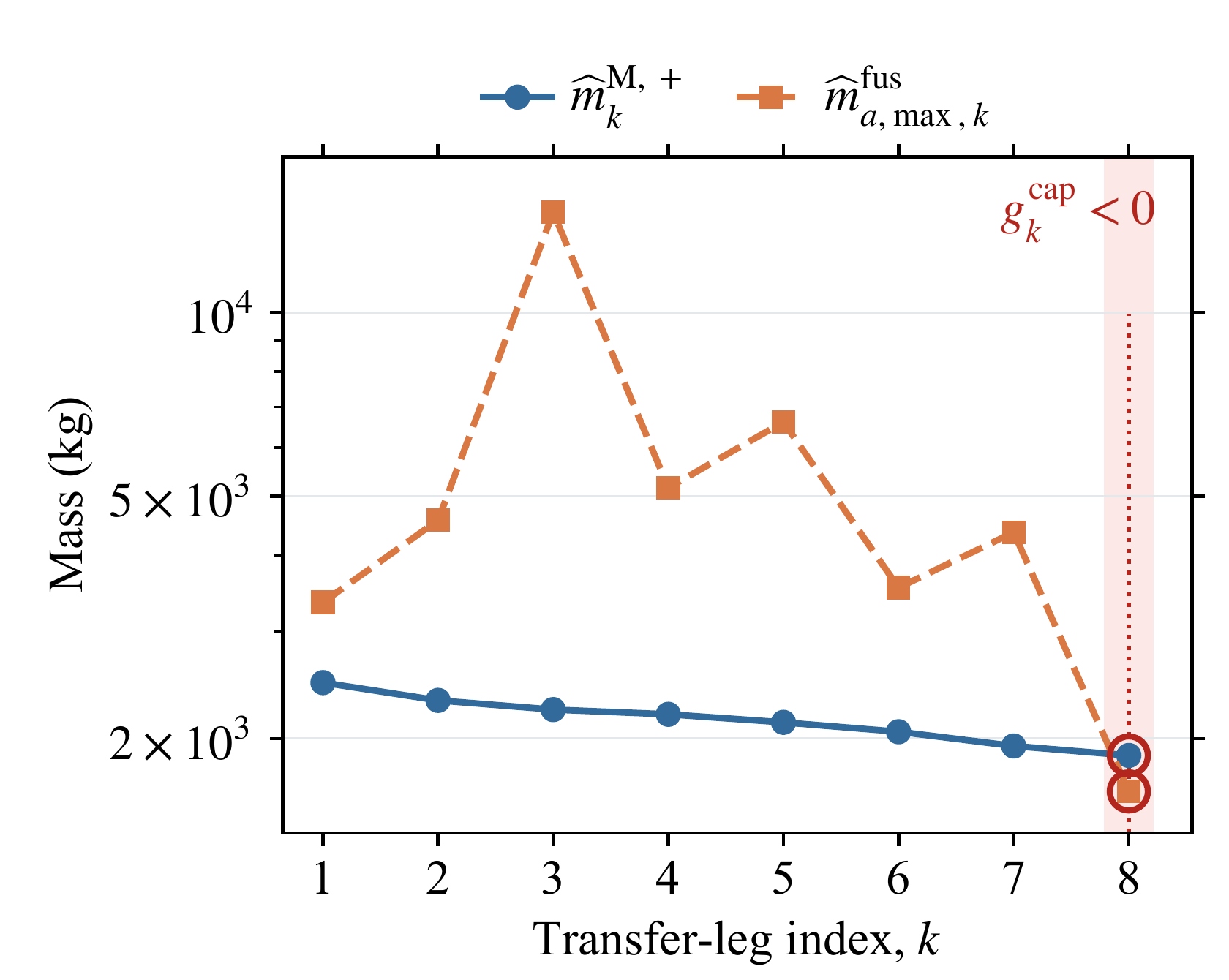}
\caption{Standard, rank 7; Sequence-Capped, rank 4}
\label{fig4.2}
\end{subfigure}
\caption{Recursive mass and predicted transfer capacity along two Lambert candidates. Red markers identify the first predicted capacity violation.}
\label{fig4}
\end{figure}

The mechanism is illustrated by the two Lambert-favored candidates marked by open symbols in Fig.~\ref{fig3}: the rank-8 Standard candidate and a sequence shared by Standard rank 7 and Sequence-Capped rank 4. Neither appears in the corresponding resource-filtered candidate set. Figure~\ref{fig4} traces their recursive departure masses $\widehat{m}_k^{\mathrm{M},+}$ against the fused capacities $\widehat{m}_{a,\max,k}^{\mathrm{fus}}$ at the search-generated encounter epochs. Each margin therefore combines the fused capacity threshold with a departure mass propagated by the propellant model across preceding legs; the highlighted transfer is the first predicted negative margin in Eq.~\eqref{eq28}.

For the rank-8 Standard candidate [Fig.~\ref{fig4}(\subref{fig4.1})], the transfer to the sixth asteroid, $k=5$, began at 2126.29~kg against a predicted capacity of 1865.85~kg. For the shared candidate [Fig.~\ref{fig4}(\subref{fig4.2})], the corresponding masses were 1876.42 and 1637.92~kg on the transfer to the ninth asteroid, $k=8$. Earlier transfers in both sequences passed the capacity test.

Both candidates remain competitive under the impulsive ranking metric but develop negative predicted capacity margins and are flagged as predicted low-thrust infeasible by the MANN resource criterion. Neither subsequently yields an accepted FT trajectory. These examples show how recursive mass propagation exposes a later-leg capacity conflict that is invisible to Lambert ranking. Within the tested fixed-time candidate population, every Lambert-selected sequence that was later not accepted by SCP had already been flagged by the MANN resource criterion before final beam allocation.

After encounter-time optimization, the three depth-synchronous Lambert searches attain a best terminal mass of 1579.559~kg, compared with 1543.352~kg for the filtered searches (Table~\ref{tab8}). Filtering retains more accepted VT entries in Standard and Sequence-Capped search, while Lambert attains the higher best mass. Thus, the gains in candidate retention at the search-generated epochs do not translate into a higher best terminal mass after timing optimization.

\subsection{Maximum-Visit Case}
\label{sec5.3}

The maximum-visit case examines whether resource filtering increases the number of asteroid encounters supported by low-thrust refinement at the search-generated epochs. The mission horizon is ten years, the beam width is $B=50$, and Time-Window search is evaluated with $\varepsilon=20$, 50, and 80 days. The event grid and transfer-duration bounds are shared with the fixed-visit case and are given in the Supplementary Material. At each evaluated search depth, up to ten ranked candidates are refined; the objective first maximizes the SCP-validated visit count and then terminal mass at that count.

Table~\ref{tab9} reports the maximum SCP-validated visit count $N_{\max}$ for each method and refinement mode. The accepted-entry count and best terminal mass in each block refer to that block's $N_{\max}$. FT assesses the encounters realizable at the search-generated epochs; VT additionally permits timing optimization. Terminal mass is a secondary objective and is compared only at equal visit count.
\begin{table}[!h]
\caption{SCP-validated visit counts and terminal masses for the maximum-visit case.}
\label{tab9}
\centering
\begin{tabular}{llrrrrrr}
\hline\hline
 & & \multicolumn{3}{c}{FT} & \multicolumn{3}{c}{VT} \\
\cline{3-5}\cline{6-8}
Architecture & Method & $N_{\max}$ & Valid & Best $m_f$, kg & $N_{\max}$ & Valid & Best $m_f$, kg \\
\hline
Standard & Lambert & 14 & 1/10 & 1298.163 & 15 & 10/10 & 1327.940 \\
Standard & MANN & 15 & 6/10 & 1006.855 & 15 & 8/10 & 1144.232 \\
Sequence-Capped & Lambert & 14 & 1/10 & 1298.163 & 15 & 10/10 & 1334.495 \\
Sequence-Capped & MANN & 15 & 6/10 & 1006.855 & 15 & 8/10 & 1144.232 \\
Round-Robin & Lambert & 13 & 1/10 & 1385.601 & 15 & 10/10 & 1325.244 \\
Round-Robin & MANN & 15 & 2/10 & 789.649 & 15 & 4/10 & 1064.983 \\
\hline
Time-Window (20 d) & Lambert & 3 & 3/10 & 2447.120 & 4 & 1/10 & 2055.290 \\
Time-Window (20 d) & MANN & 5 & 9/10 & 2157.281 & 5 & 9/10 & 2344.092 \\
Time-Window (50 d) & Lambert & 3 & 3/10 & 2447.120 & 5 & 1/10 & 1817.926 \\
Time-Window (50 d) & MANN & 4 & 10/10 & 2362.984 & 4 & 10/10 & 2425.794 \\
Time-Window (80 d) & Lambert & 3 & 3/10 & 2447.120 & 4 & 1/10 & 2055.290 \\
Time-Window (80 d) & MANN & 4 & 10/10 & 2396.109 & 4 & 10/10 & 2457.500 \\
\hline\hline
\end{tabular}
\end{table}

At the search-generated epochs, resource filtering enables longer SCP-validated tours. All three depth-synchronous searches reach the 15-visit level, but filtering increases the maximum SCP-validated visit count from 14 to 15 in Standard and Sequence-Capped search and from 13 to 15 in Round-Robin search. At the 15-visit level, none of the ten highest-ranked Lambert candidates yields an accepted FT trajectory, whereas the filtered searches yield six, six, and two accepted entries, respectively. The gain therefore arises from retaining sequences whose low-thrust resource evolution remains compatible with the prescribed encounter epochs, rather than simply expanding the discrete search to greater depth.

\begin{figure}[!h]
\centering
\begin{subfigure}[t]{0.49\linewidth}
\centering
\includegraphics[width=\linewidth]{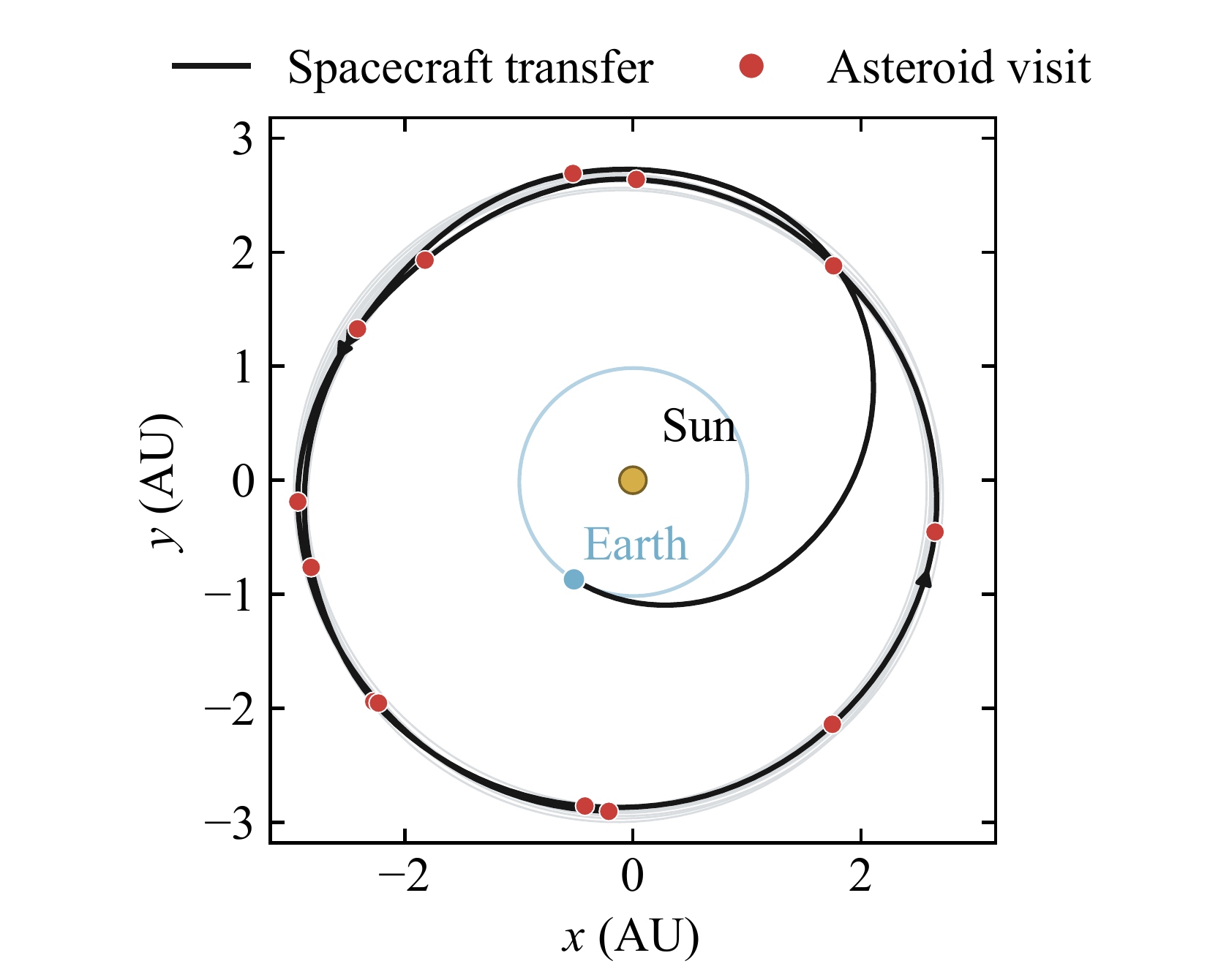}
\caption{Lambert, $N=13$}
\label{fig5.1}
\end{subfigure}\hfill
\begin{subfigure}[t]{0.49\linewidth}
\centering
\includegraphics[width=\linewidth]{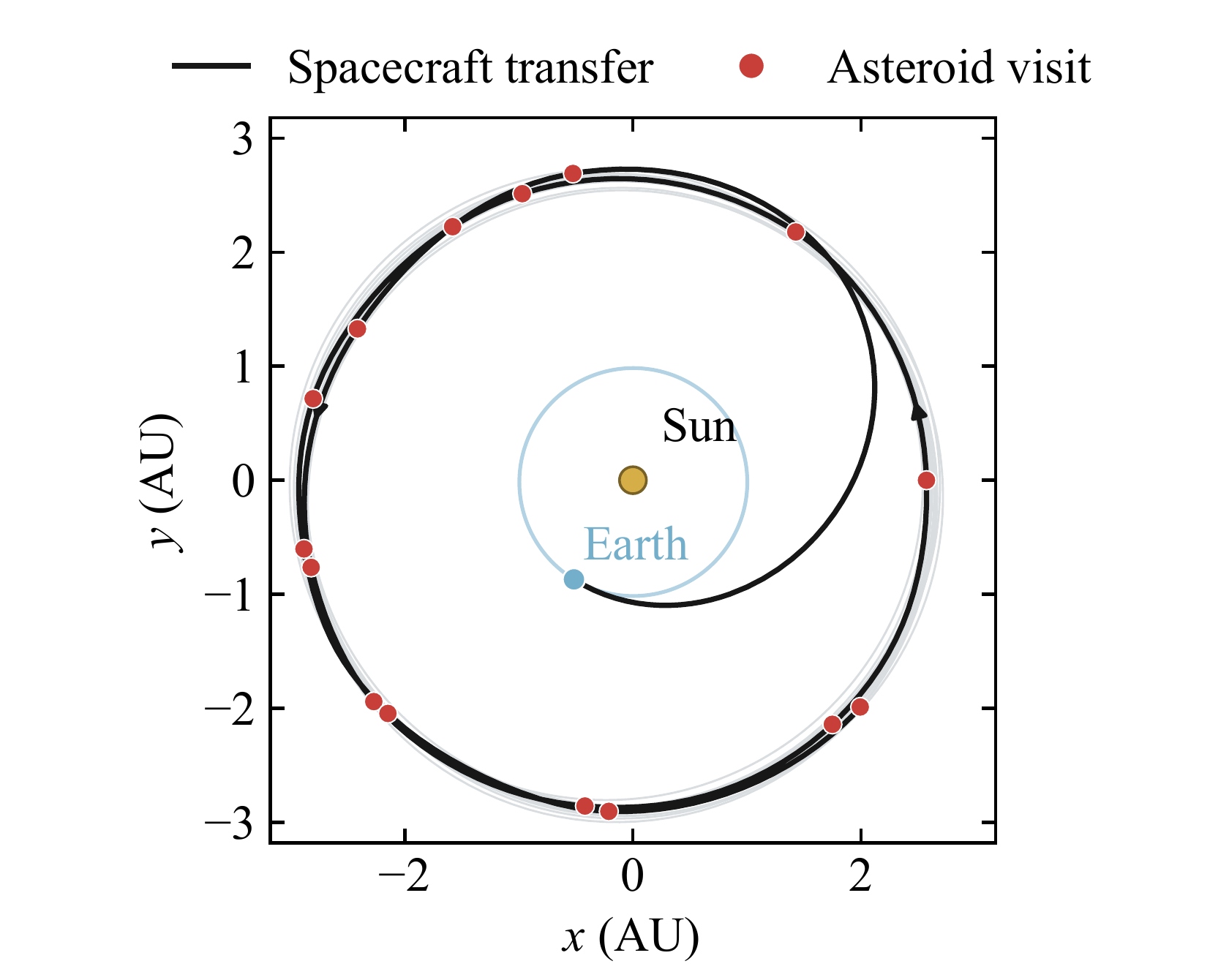}
\caption{MANN, $N=15$}
\label{fig5.2}
\end{subfigure}
\caption{Heliocentric projections of representative fixed-time low-thrust trajectories from Round-Robin beam search.}
\label{fig5}
\end{figure}

Figure~\ref{fig5} illustrates the Round-Robin result through the accepted FT trajectory with the highest terminal mass at each method's maximum SCP-validated visit count. The Lambert trajectory visits 13 asteroids, whereas the resource-filtered trajectory visits all 15 without adjusting the search epochs.

The same qualitative fixed-time trend appears in Time-Window search, where filtering increases the maximum SCP-validated visit count for all three tested window widths, although the achievable tour lengths remain lower than those of the depth-synchronous architectures.

With encounter-time optimization, all three depth-synchronous Lambert searches attain 15 validated visits, matching the filtered searches, and yield higher best terminal masses at that visit count. The FT advantage therefore lies in retaining candidates that support 15 visits at the original search epochs. Table~\ref{tab9} gives the window-dependent VT results.

\subsection{Deployment--Collection Case}
\label{sec5.4}

The deployment--collection case couples transfer feasibility to returned mined mass. Nine asteroids are selected for deployment and collection before Earth return, and the objective is the returned mined mass $M_{\mathrm{ret}}$. The mission spans approximately 15 years. Each 40-kg miner produces 10~kg per year and remains deployed for at least 365 days. Beam widths are $B=50$ and 100. Earth departure and return excess speeds are bounded by 6~km/s. Both search variants use the same SCP-generated Earth-departure seeds. The epoch windows, event grid, and transfer-duration bounds are specified in the Supplementary Material.

Standard, Sequence-Capped, and Round-Robin architectures are compared. Time-Window search is omitted because its asynchronous expansion is not directly comparable to the paired forward--reverse construction used in this case. Deployment branches use the forward resource filter of Sec.~\ref{sec4.2}. Because the physical payload and mass history are unavailable until the deployment and collection branches are paired, collection orders are constructed backward with representative spacecraft masses decreasing from 1300 to 1200~kg over the reverse collection levels; the exact schedule is given in the Supplementary Material. The resulting capacity screen is a construction-level proxy rather than a feasibility test at the actual mission mass. After pairing, the complete mission is replayed forward in physical time using predicted propellant consumption and the actual deployment and collection mass changes, followed by Earth-return closure and full-mission SCP refinement.

\begin{figure}[!h]
\centering
\includegraphics[width=0.85\linewidth]{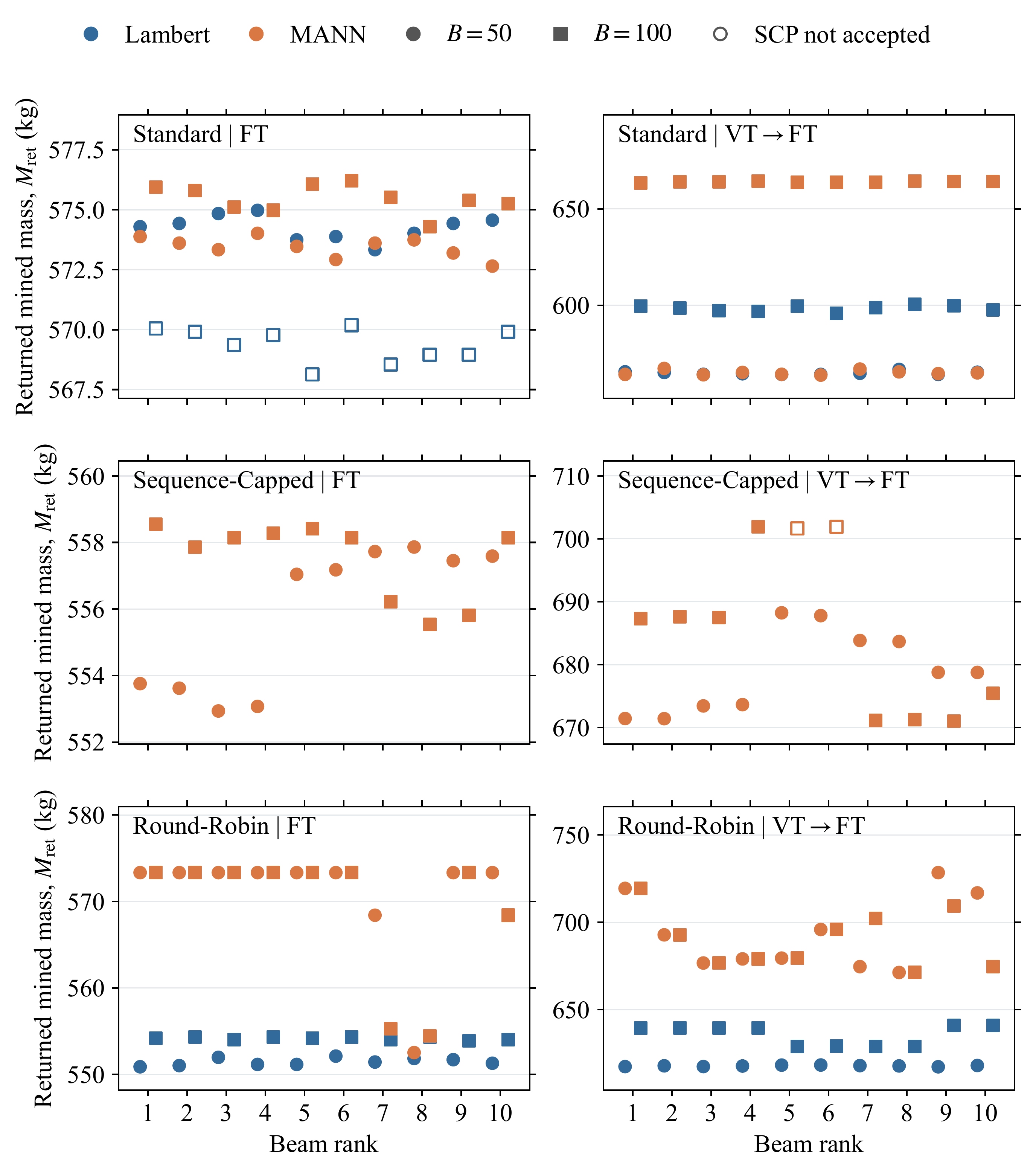}
\caption{Returned mined mass by beam rank. Color distinguishes search method, shape beam width, and fill SCP acceptance across FT and VT$\rightarrow$FT.}
\label{fig6}
\end{figure}

Timing changes alter both mining duration and the cost of transporting the collected payload. Timing-adjusted epochs are refined in an additional fixed-time pass, denoted VT$\rightarrow$FT. Table~\ref{tab10} reports the original FT and timing-adjusted results, including the number $N_{\mathrm{term}}$ of terminal candidates, the accepted full-mission SCP trajectories among the ten evaluated candidates, and the best returned mined mass among those trajectories.
Figure~\ref{fig6} compares returned mined mass before and after timing adjustment. Open markers denote sequence candidates for which SCP did not obtain a full-mission trajectory satisfying the dynamical and mass-consistency requirements. No Sequence-Capped Lambert series appears because that search produced no terminal candidates.

\begin{table}[!h]
\caption{Deployment--collection candidate availability and returned mass; MANN uses forward filtering, proxy-based reverse construction, and joined forward replay.}
\label{tab10}
\centering
\begin{tabular}{lrllrrrr}
\hline\hline
 & & & & \multicolumn{2}{c}{FT} & \multicolumn{2}{c}{VT$\rightarrow$FT} \\
\cline{5-6}\cline{7-8}
Architecture & $B$ & Method & $N_{\mathrm{term}}$ & Valid & Best $M_{\mathrm{ret}}$, kg & Valid & Best $M_{\mathrm{ret}}$, kg \\
\hline
Standard & 50 & Lambert & 50 & 10/10 & 574.976 & 10/10 & 566.804 \\
Standard & 50 & MANN & 50 & 10/10 & 574.018 & 10/10 & 567.298 \\
Standard & 100 & Lambert & 100 & 0/10 & --- & 10/10 & 600.479 \\
Standard & 100 & MANN & 100 & 10/10 & 576.208 & 10/10 & 664.397 \\
\hline
Sequence-Capped & 50 & Lambert & 0 & --- & --- & --- & --- \\
Sequence-Capped & 50 & MANN & 14 & 10/10 & 557.864 & 10/10 & 688.239 \\
Sequence-Capped & 100 & Lambert & 0 & --- & --- & --- & --- \\
Sequence-Capped & 100 & MANN & 46 & 10/10 & 558.549 & 8/10 & 701.917 \\
\hline
Round-Robin & 50 & Lambert & 50 & 10/10 & 552.115 & 10/10 & 618.326 \\
Round-Robin & 50 & MANN & 30 & 10/10 & 573.333 & 10/10 & 728.406 \\
Round-Robin & 100 & Lambert & 100 & 10/10 & 554.305 & 10/10 & 641.098 \\
Round-Robin & 100 & MANN & 71 & 10/10 & 573.333 & 10/10 & 719.394 \\
\hline\hline
\end{tabular}
\end{table}

The effect of MANN-assisted resource construction is strongly architecture dependent in the deployment--collection problem. Standard search at $B=50$ is nearly unchanged, whereas at $B=100$ it changes from no accepted FT trajectory to ten accepted entries. Sequence-Capped produces terminal candidates and accepted full missions only with MANN-assisted construction. By contrast, Round-Robin already produces accepted trajectories with either method, and its principal change is higher returned mined mass across most evaluated ranks [Fig.~\ref{fig6}].

\begin{figure}[!h]
\centering
\begin{subfigure}[t]{0.49\linewidth}
\centering
\includegraphics[width=\linewidth]{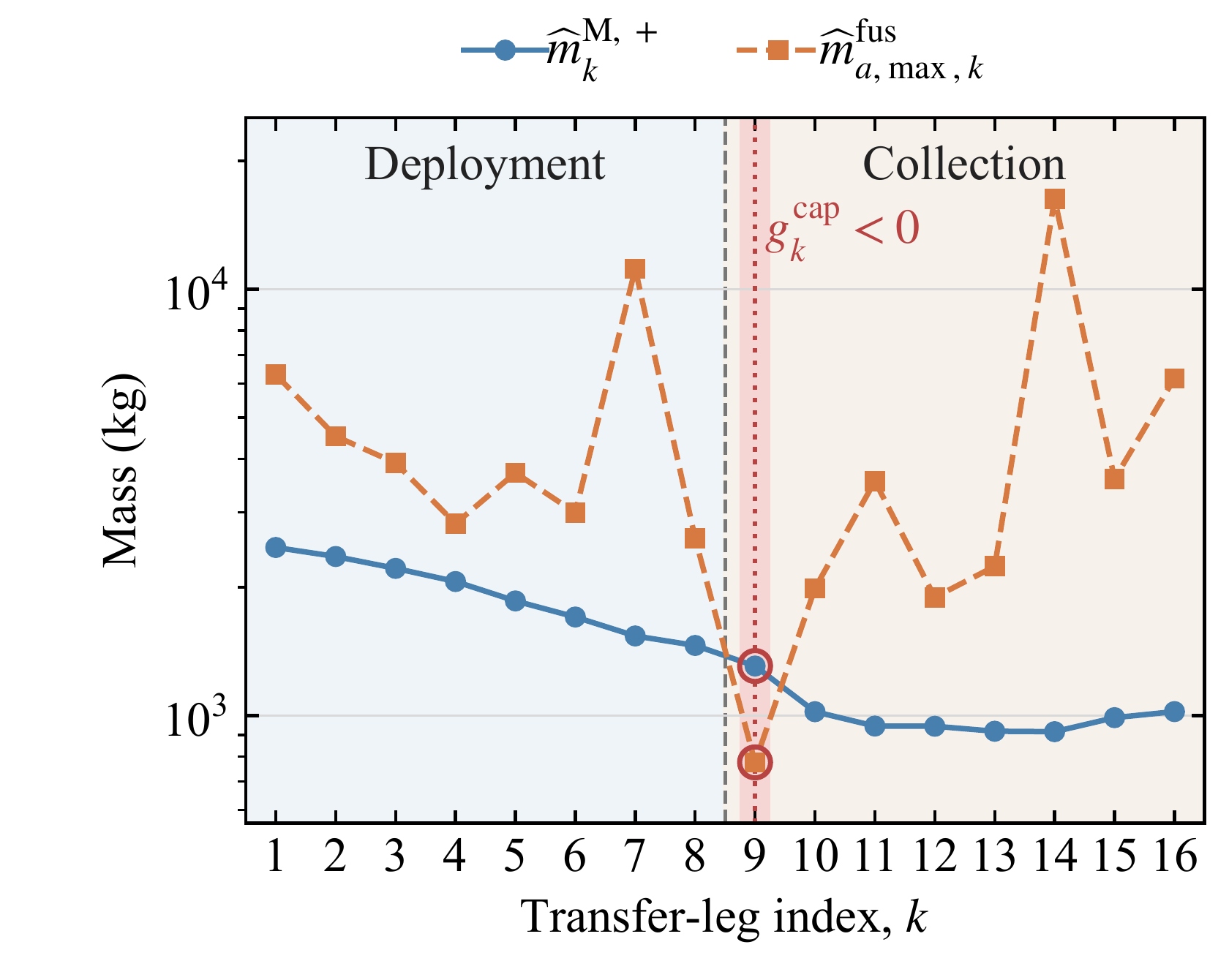}
\caption{Lambert candidate}
\label{fig7.1}
\end{subfigure}\hfill
\begin{subfigure}[t]{0.49\linewidth}
\centering
\includegraphics[width=\linewidth]{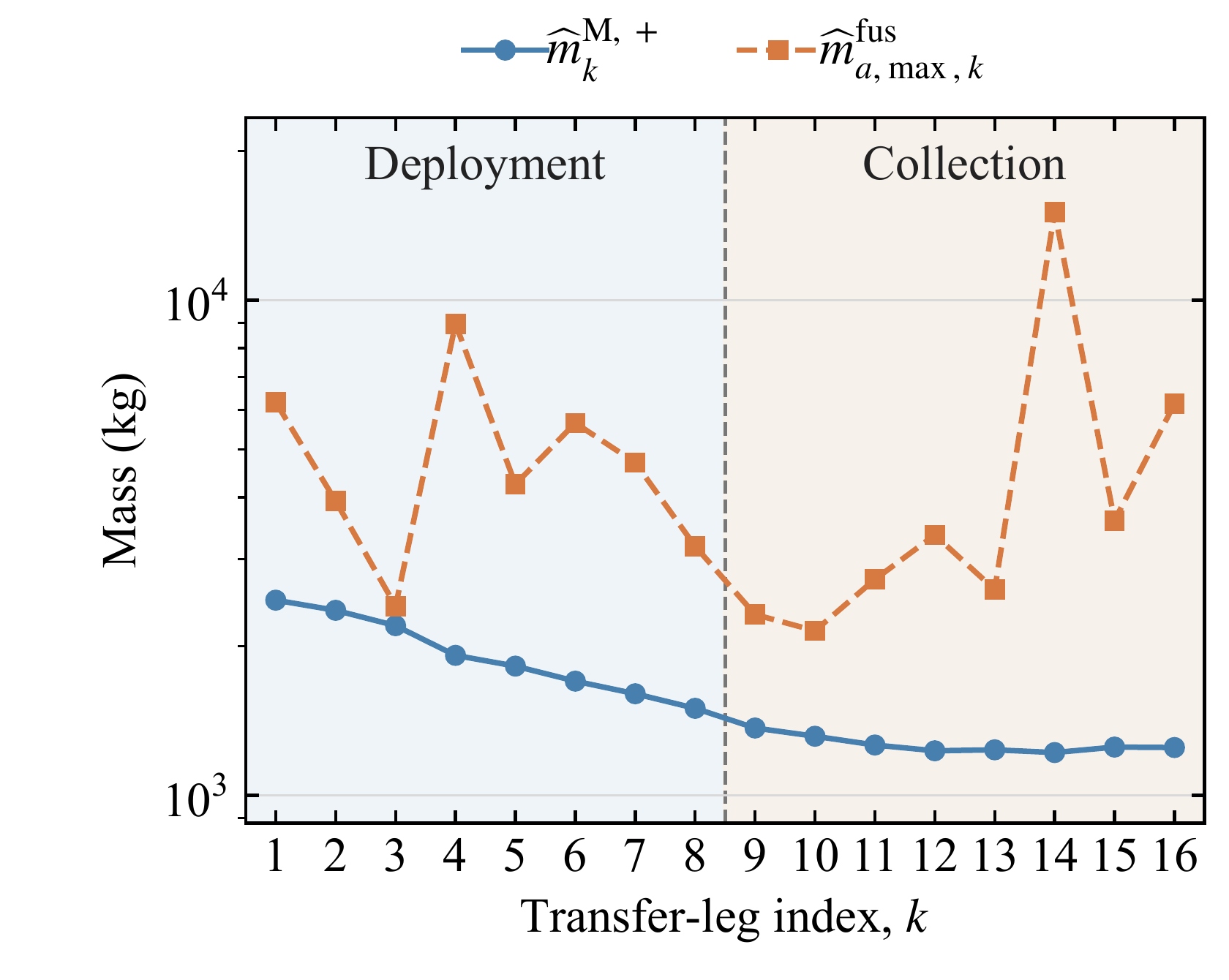}
\caption{MANN-assisted candidate}
\label{fig7.2}
\end{subfigure}
\caption{Recursive mass and predicted transfer capacity for two deployment--collection candidates from Standard beam search at $B=100$.}
\label{fig7}
\end{figure}

After the deployment and collection orders are joined, Fig.~\ref{fig7} provides a transfer-level explanation for this behavior by contrasting fixed-epoch mass--capacity profiles of the rank-1 Lambert and rank-8 MANN-assisted Standard candidates at $B=100$. Each margin compares the fused transfer-capacity estimate with the departure mass obtained by forward propagation using the propellant predictions. The Lambert sequence first reaches a negative predicted capacity margin during collection, whereas the MANN-assisted sequence retains positive margins. The plotted index counts the sixteen inter-asteroid transfers, omitting the Earth boundary legs and the intervening stay.

The Lambert candidate reaches a negative predicted capacity margin on the first inter-asteroid collection transfer, from asteroid 3896 to 19434. Its recursive departure mass is 1306.37~kg, against a capacity of 776.31~kg. The MANN-assisted candidate maintains positive margins throughout, with a minimum of 209.97~kg. The replay illustrates how the combined capacity and recursive-mass criterion distinguishes the two candidates.

Timing adjustment increases the returned-mass difference in several comparisons (Table~\ref{tab10}). The largest observed gain occurs for Round-Robin at $B=50$, where the best returned mined mass increases from 618.326 to 728.406~kg, or 17.8\%. Within the tested beam widths, increasing $B$ from 50 to 100 does not consistently improve the best returned mined mass. These results suggest that beam width alone does not determine final deployment--collection performance, because the retained encounter sequences and epochs also affect both mining duration and low-thrust transport.

\begin{figure}[!h]
\centering
\begin{subfigure}[t]{0.49\linewidth}
\centering
\includegraphics[width=\linewidth]{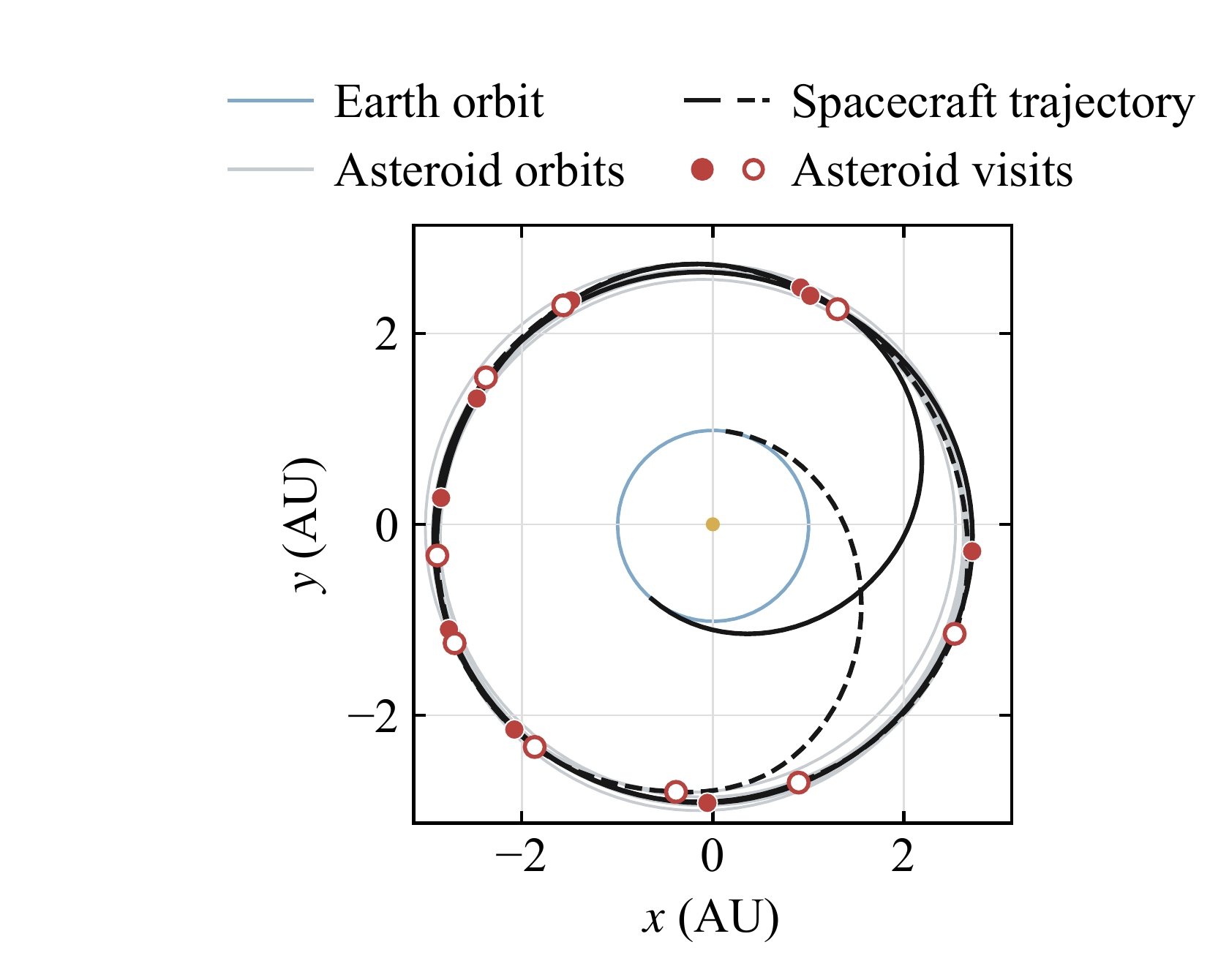}
\caption{Heliocentric $x$--$y$ projection}
\label{fig8.1}
\end{subfigure}\hfill
\begin{subfigure}[t]{0.49\linewidth}
\centering
\includegraphics[width=\linewidth]{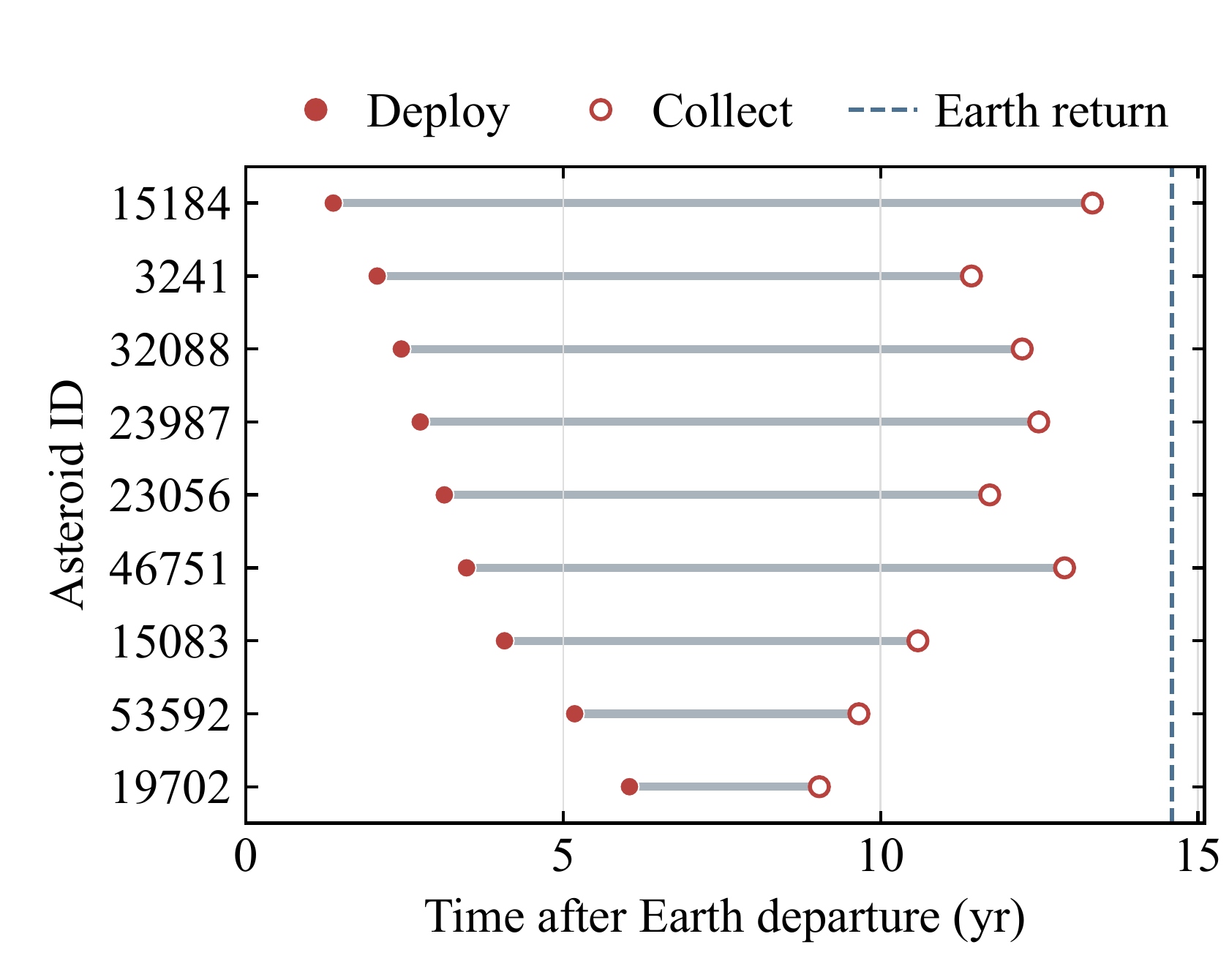}
\caption{Deployment--collection intervals}
\label{fig8.2}
\end{subfigure}
\caption{Heliocentric trajectory projection and mining intervals for the accepted deployment--collection mission with the highest returned mined mass.}
\label{fig8}
\end{figure}

Figure~\ref{fig8} links the highest-yield accepted trajectory to its deployment and collection intervals. All miners were deployed within 6.044~yr, collection began at 9.036~yr, and Earth return occurred at 14.591~yr. Mining intervals ranged from 2.992 to 11.959~yr. At 10~kg per year per miner, their combined production yielded 728.406~kg of returned mined mass.

\section{Conclusion}
\label{sec6}
This study develops a Lambert-ranked beam-search method that combines direct continuous prediction of low-thrust transfer capacity and mass-dependent propellant consumption for multi-asteroid rendezvous optimization. The MIM and propellant models achieve independent-test MAEs of 40.92~kg for maximum admissible initial mass after fusion and 1.1713~kg for propellant consumption. The capacity prediction screens candidate extensions, and the propellant prediction propagates branch mass; together, they improve the population eligible for SCP refinement. Across the fixed-visit, maximum-visit, and deployment--collection cases, filtering increases accepted fixed-time trajectories, raises the maximum SCP-validated visit count to 15, and produces a 17.8\% improvement in the highest returned mined mass. These results demonstrate that learned transfer-level resource information can improve finite-width search for low-thrust multi-asteroid missions.

\appendix
\setcounter{table}{0}
\setcounter{figure}{0}
\renewcommand{\thetable}{\thesection.\arabic{table}}
\renewcommand{\thefigure}{\thesection.\arabic{figure}}
\renewcommand{\theHtable}{appendix.\Alph{section}.\arabic{table}}
\renewcommand{\theHfigure}{appendix.\Alph{section}.\arabic{figure}}

\section{Sensitivity Analysis of the MIM Model}
\label{sec7.1}

Table~\ref{tabA1} compares MIM-model inputs, coordinates, and development-set sizes. F0 contains the reduced boundary-state features and transfer duration. F1 adds $T/P_a$; F2 and F3 then add the sum and the separate values of $\Delta v_1$ and $\Delta v_2$, respectively. F4 and F5 add the analytical mass estimates $(m_a^{\mathrm{g}},m_b^{\mathrm{g}})$ to F2 and F3. Each of F6--F11 adds forward- and backward-propagation residuals to its counterpart F0--F5; F11 is the selected input in Eq.~\eqref{eq18}. The coordinate abbreviations Cart., Cyl., Sph., Class., Equin., and Mod. denote Cartesian, cylindrical, spherical, classical, equinoctial, and modified orbital elements. GNLL uses normalized outputs, MAEs are in kilograms, and 250k denotes 250,000 development records. Complete feature and coordinate definitions are given in the Supplementary Material.


\begin{table}[!h]
\caption{Capacity-model sensitivity comparisons.}
\label{tabA1}
\centering
\begin{tabular}{lrrrlrrr}
\hline\hline
Case & GNLL & $\mathrm{MAE}_a$, kg & $\mathrm{MAE}_b$, kg & Case & GNLL & $\mathrm{MAE}_a$, kg & $\mathrm{MAE}_b$, kg\\
\midrule
\multicolumn{8}{@{}l}{\textit{Input features: Spherical, 250k, 6/64}}\\*
F0 & -2.8595 & 993.36 & 993.25 & F6 & -5.0207 & 175.01 & 174.74\\
F1 & -2.7499 & 1060.82 & 1060.96 & F7 & -5.6384 & 136.28 & 135.16\\
F2 & -3.5768 & 489.68 & 489.53 & F8 & -5.6185 & 120.42 & 120.27\\
F3 & -4.5398 & 222.59 & 222.53 & F9 & -5.4723 & 88.37 & 87.81\\
F4 & -3.6941 & 432.38 & 432.94 & F10 & -5.5132 & 130.00 & 130.23\\
F5 & -4.3732 & 260.64 & 260.14 & F11 & -6.2730 & 62.88 & 63.28\\
\midrule
\multicolumn{8}{@{}l}{\textit{Coordinates: F11, 250k, 6/64}}\\*
Cart. & -6.3081 & 68.08 & 67.06 & Class. & -4.9458 & 163.68 & 162.99\\
Cyl. & -6.1390 & 74.79 & 74.92 & Equin. & -5.7585 & 102.11 & 102.74\\
Sph. & -6.2730 & 62.88 & 63.28 & Mod. & -5.6960 & 104.33 & 103.07\\
\midrule
\multicolumn{8}{@{}l}{\textit{Development records: Spherical, F11, 6/64}}\\*
50,000 & -5.5738 & 88.01 & 87.85 & 250,000 & -6.2730 & 62.88 & 63.28\\
100,000 & -5.0333 & 132.84 & 128.61 & 1M & -7.0191 & 41.45 & 40.97\\
\hline\hline
\end{tabular}

\end{table}

Table~\ref{tabA2} compares MANN blocks and latent width using spherical F11 inputs and 250,000 development records; all MAEs are in kilograms.

\begin{table}[!h]
\caption{Capacity-model MANN configuration comparisons.}
\label{tabA2}
\centering
\begin{tabular}{rrrrrrrrrr}
\hline\hline
Blocks & Width & GNLL & $\mathrm{MAE}_{a}$ & $\mathrm{MAE}_b$ & Blocks & Width & GNLL & $\mathrm{MAE}_{a}$ & $\mathrm{MAE}_b$ \\
\hline
4 & 32 & -5.2137 & 156.26 & 154.33 & 8 & 32 & -5.9909 & 86.64 & 84.44 \\
4 & 64 & -5.8436 & 96.68 & 96.63 & 8 & 64 & -6.0481 & 66.86 & 66.59 \\
4 & 128 & -6.1772 & 80.45 & 80.50 & 8 & 128 & -5.4749 & 92.29 & 96.53 \\
4 & 256 & -5.7262 & 82.61 & 83.91 & 8 & 256 & -1.3472 & 2290.94 & 2189.11 \\
\hline
6 & 32 & -5.7328 & 105.40 & 106.47 & 10 & 32 & -6.0621 & 75.07 & 76.11 \\
6 & 64 & -6.2730 & 62.88 & 63.28 & 10 & 64 & -5.3004 & 96.90 & 95.32 \\
6 & 128 & -5.3848 & 85.34 & 80.53 & 10 & 128 & -5.4941 & 93.41 & 92.18 \\
6 & 256 & -5.0630 & 77.89 & 77.49 & 10 & 256 & -1.3472 & 2291.14 & 2193.02 \\
\hline\hline
\end{tabular}
\end{table}

\setcounter{table}{0}
\setcounter{figure}{0}
\section{Sensitivity Analysis of the Propellant Model}
\label{sec7.2}

Table~\ref{tabB1} compares propellant-model inputs, coordinates, and development-set sizes. F0 contains $(\boldsymbol{\chi},\Delta v_1,\Delta v_2,m_a,m_b^{\mathrm{L}})$. F1 adds $m_{a,\max}$ and F2 adds the mass ratio $\rho$. F3 adds $(m_{a,\max},q)$, whereas F4 adds $(\rho,q)$ and corresponds to the selected input in Eq.~\eqref{eq22}. The feature and coordinate scans use $6/64$, while the dataset-size scan uses $10/64$. GNLL is evaluated on normalized outputs, MAEs are in kilograms, and 250k denotes 250,000 development records. The Supplementary Material gives the complete feature definitions and uncertainty statistics.

\begin{table}[!h]
\caption{Propellant-model sensitivity comparisons.}
\label{tabB1}
\centering
\begin{tabular}{lrrrlrrr}
\hline\hline
Case & GNLL & $\mathrm{MAE}_{\mathrm{p}}$, kg & $\mathrm{MAE}_b$, kg & Case & GNLL & $\mathrm{MAE}_{\mathrm{p}}$, kg & $\mathrm{MAE}_b$, kg\\
\midrule
\multicolumn{8}{@{}l}{\textit{Input features: Spherical, 250k, 6/64}}\\*
F0 & -6.0811 & 5.5356 & 6.9702 & F3 & -6.2489 & 3.4025 & 5.5311\\
F1 & -6.1667 & 3.7367 & 5.8917 & F4 & -6.2943 & 2.8181 & 6.3455\\
F2 & -6.2283 & 2.8533 & 7.0166 &  &  &  & \\
\midrule
\multicolumn{8}{@{}l}{\textit{Coordinates: F4, 250k, 6/64}}\\*
Cart. & -6.2494 & 2.9049 & 6.5977 & Class. & -6.0443 & 3.7294 & 7.5227\\
Cyl. & -6.2253 & 2.8595 & 6.8761 & Equin. & -6.2110 & 2.8530 & 7.1921\\
Sph. & -6.2943 & 2.8181 & 6.3455 & Mod. & -6.1283 & 3.4444 & 7.0410\\
\midrule
\multicolumn{8}{@{}l}{\textit{Development records: Spherical, F4, 10/64}}\\*
50,000 & -5.4899 & 2.6785 & 11.3898 & 250,000 & -6.5948 & 1.9314 & 6.3467\\
100,000 & -6.2231 & 2.1866 & 8.0399 & 1M & -7.3431 & 1.1713 & 3.3569\\
\hline\hline
\end{tabular}

\end{table}

Table~\ref{tabB2} uses spherical F4 inputs and 250,000 development records.

\begin{table}[!h]
\caption{Propellant-model MANN configuration comparisons.}
\label{tabB2}
\centering
\begin{tabular}{rrrrrrrrrr}
\hline\hline
Blocks & Width & GNLL & $\mathrm{MAE}_{\mathrm{p}}$ & $\mathrm{MAE}_b$ & Blocks & Width & GNLL & $\mathrm{MAE}_{\mathrm{p}}$ & $\mathrm{MAE}_b$ \\
\hline
4 & 32 & -5.7084 & 5.9060 & 10.7411 & 8 & 32 & -5.9998 & 3.6544 & 7.9505 \\
4 & 64 & -6.0115 & 3.8382 & 7.2289 & 8 & 64 & -6.4080 & 2.3448 & 6.3729 \\
4 & 128 & -6.2560 & 3.0018 & 5.7601 & 8 & 128 & -6.5234 & 2.4432 & 5.4257 \\
4 & 256 & -4.0224 & 55.9547 & 23.4538 & 8 & 256 & -1.3766 & 105.0406 & 1571.1572 \\
\hline
6 & 32 & -5.8649 & 4.3537 & 9.3993 & 10 & 32 & -6.0512 & 3.2791 & 9.5363 \\
6 & 64 & -6.2943 & 2.8181 & 6.3455 & 10 & 64 & -6.5948 & 1.9314 & 6.3467 \\
6 & 128 & -6.4920 & 2.4350 & 5.3263 & 10 & 128 & -4.6064 & 7.2267 & 16.1694 \\
6 & 256 & -5.3349 & 35.4623 & 9.3759 & 10 & 256 & -1.3766 & 104.9698 & 1570.7510 \\
\hline\hline
\end{tabular}
\end{table}

\setcounter{table}{0}
\setcounter{figure}{0}
\section{Asteroid Candidate Set}
\label{sec7.3}

Table~\ref{tabC1} gives the heliocentric orbital elements of the 15 GTOC12 catalogue targets used in all three cases. The common catalogue epoch is modified Julian date (MJD) 64328; $\Omega$, $\omega$, and $M$ denote longitude of ascending node, argument of perihelion, and mean anomaly, respectively.

\begin{table}[!h]
\caption{Orbital elements of the shared asteroid candidate set.}
\label{tabC1}
\centering
\begin{tabular}{rrrrrrr}
\hline\hline
ID & $a$, AU & $e$ & $i$, deg & $\Omega$, deg & $\omega$, deg & $M$, deg\\
\hline
2032 & 2.759 & 0.0853 & 2.77 & 223.10 & 201.36 & 216.3995\\
3241 & 2.793 & 0.0385 & 5.91 & 81.87 & 315.38 & 253.5117\\
3896 & 2.786 & 0.0782 & 5.27 & 107.77 & 275.66 & 262.9474\\
5707 & 2.781 & 0.0704 & 1.57 & 40.97 & 347.34 & 261.3867\\
15083 & 2.777 & 0.0614 & 5.58 & 86.22 & 330.41 & 231.2066\\
15184 & 2.777 & 0.0855 & 1.62 & 73.91 & 295.49 & 276.8766\\
19434 & 2.799 & 0.0667 & 4.46 & 19.29 & 27.88 & 245.4491\\
19702 & 2.786 & 0.0689 & 4.26 & 184.72 & 219.22 & 244.4074\\
23056 & 2.799 & 0.0682 & 5.18 & 60.08 & 340.36 & 250.4891\\
23987 & 2.788 & 0.0890 & 5.13 & 55.41 & 5.59 & 223.4749\\
32088 & 2.788 & 0.0639 & 5.24 & 58.00 & 352.61 & 236.8349\\
37818 & 2.766 & 0.0668 & 2.06 & 107.67 & 302.77 & 231.0051\\
46418 & 2.777 & 0.0759 & 4.76 & 173.38 & 225.12 & 248.9866\\
46751 & 2.794 & 0.0714 & 5.34 & 67.45 & 330.05 & 253.9184\\
53592 & 2.810 & 0.0605 & 4.54 & 175.63 & 243.47 & 237.7185\\
\hline\hline
\end{tabular}
\end{table}

\subsection*{Acknowledgements}
This work was supported by the Civil Space Technology Advance Research
Project (D010103) and the National Natural Science Foundation of China
(No. 124B2049). The authors thank Dong Qiao for helpful discussions on
neural-network modeling. Xingyu Zhou is
grateful for the financial support provided by the China Scholarship
Council (No. 202406030186).

\subsection*{Declaration of competing interest}

The authors have no competing interests to declare that are relevant to
the content of this article.

\section*{References}
\bibliographystyle{astrobib}
\bibliography{sample,mann_references}

@article{Zhu2009MultiAsteroid,
  author  = {Zhu, Kaijian and Jiang, Fanghua and Li, Junfeng and Baoyin, Hexi},
  title   = {Trajectory Optimization of Multi-Asteroids Exploration with Low Thrust},
  journal = {Transactions of the Japan Society for Aeronautical and Space Sciences},
  volume  = {52}, number = {175}, pages = {47--54}, year = {2009},
  doi     = {10.2322/tjsass.52.47}
}

@article{Tang2016Adjoint,
  author  = {Jiang, Fanghua and Tang, Gao},
  title   = {Systematic Low-Thrust Trajectory Optimization for a Multi-Rendezvous Mission using Adjoint Scaling},
  journal = {Astrophysics and Space Science},
  volume  = {361}, number = {4}, pages = {117}, year = {2016},
  doi     = {10.1007/s10509-016-2704-5}
}

@article{DiCarlo2018MainBelt,
  author  = {Di Carlo, Marco and Vasile, Massimiliano and Dunlop, Michael W.},
  title   = {Low-thrust tour of the main belt asteroids},
  journal = {Advances in Space Research},
  volume  = {62}, number = {8}, pages = {2026--2045}, year = {2018},
  doi     = {10.1016/j.asr.2017.12.033}
}

@article{Zhang2025GTOC12Design,
  author  = {Zhang, Zhong and Zhang, Nan and Guo, Xiang and Wu, Di and Xie, Xuan and Yang, Jia and Jiang, Fanghua and Baoyin, Hexi},
  title   = {Sustainable Asteroid Mining: On the Design of {GTOC12} Problem and Summary of Results},
  journal = {Astrodynamics}, volume = {9}, number = {1}, pages = {3--17}, year = {2025},
  doi     = {10.1007/s42064-024-0199-3}
}

@article{Betts1998Survey,
  author  = {Betts, John T.},
  title   = {Survey of Numerical Methods for Trajectory Optimization},
  journal = {Journal of Guidance, Control, and Dynamics},
  volume  = {21}, number = {2}, pages = {193--207}, year = {1998},
  doi     = {10.2514/2.4231}
}

@article{Fan2019MCTS,
  author  = {Fan, Zichen and Huo, Mingying and Qi, Naiming and Xu, Ye and Song, Zhiguo},
  title   = {Fast Preliminary Design of Low-Thrust Trajectories for Multi-Asteroid Exploration},
  journal = {Aerospace Science and Technology}, volume = {93}, pages = {105295}, year = {2019},
  doi     = {10.1016/j.ast.2019.07.028}
}

@article{Qi2022MCTS,
  author  = {Qi, Naiming and Fan, Zichen and Huo, Mingying and Du, Dong and Zhao, Chao},
  title   = {Fast Trajectory Generation and Asteroid Sequence Selection in Multispacecraft for Multiasteroid Exploration},
  journal = {IEEE Transactions on Cybernetics}, volume = {52}, number = {7}, pages = {6071--6082}, year = {2022},
  doi     = {10.1109/TCYB.2020.3040799}
}

@article{GTOC11Tsinghua,
  author  = {Zhang, Zhong and others},
  title   = {{GTOC} 11: Results from Tsinghua University and Shanghai Institute of Satellite Engineering},
  journal = {Acta Astronautica}, volume = {202}, pages = {819--828}, year = {2023},
  doi     = {10.1016/j.actaastro.2022.06.028}
}

@article{GTOC11Eccentric,
  author  = {Ortega Absil, J. and Serra, R. and Sumelzo Martinez, S. and Charpigny, B. and Labroquere, J. and Munoz Morales, A. and Olympio, J. and Rodriguez-Fernandez, V.},
  title   = {Design of Impulsive Asteroid Flybys and Scheduling of Time-Minimal Optimal Control Arcs for the Construction of a Dyson Ring ({GTOC} 11)},
  journal = {Acta Astronautica}, volume = {201}, pages = {94--110}, year = {2022},
  doi     = {10.1016/j.actaastro.2022.08.028}
}

@article{TheAntipodes2025,
  author  = {Armellin, Roberto and Bellome, Andrea and Fu, Xiaoyu and Holt, Harry and Parigini, Cristina and Wijayatunga, Minduli and Yarndley, Jack},
  title   = {{GTOC12}: Results from {TheAntipodes}},
  journal = {Astrodynamics}, volume = {9}, number = {1}, pages = {55--75}, year = {2025},
  doi     = {10.1007/s42064-024-0219-3}
}

@article{Izzo2015Lambert,
  author  = {Izzo, Dario},
  title   = {Revisiting {Lambert}'s Problem},
  journal = {Celestial Mechanics and Dynamical Astronomy}, volume = {121}, number = {1}, pages = {1--15}, year = {2015},
  doi     = {10.1007/s10569-014-9587-y}
}

@article{Shang2018GPR,
  author  = {Shang, Haibin and Wu, Xiaoyu and Qiao, Dong and Huang, Xiangyu},
  title   = {Parameter Estimation for Optimal Asteroid Transfer Trajectories Using Supervised Machine Learning},
  journal = {Aerospace Science and Technology}, volume = {79}, pages = {570--579}, year = {2018},
  doi     = {10.1016/j.ast.2018.06.002}
}

@article{Zhu2019MLP,
  author  = {Zhu, Yue-he and Luo, Ya-Zhong},
  title   = {Fast Evaluation of Low-Thrust Transfers via Multilayer Perceptions},
  journal = {Journal of Guidance, Control, and Dynamics}, volume = {42}, number = {12}, pages = {2627--2637}, year = {2019},
  doi     = {10.2514/1.G004080}
}

@article{Li2020DeepSurrogates,
  author  = {Li, Hai and Chen, Shuang and Izzo, Dario and Baoyin, Hexi},
  title   = {Deep Networks as Approximators of Optimal Low-Thrust and Multi-Impulse Cost in Multitarget Missions},
  journal = {Acta Astronautica}, volume = {166}, pages = {469--481}, year = {2020},
  doi     = {10.1016/j.actaastro.2019.09.023}
}

@article{Dempster2021OnlineDNN,
  author  = {Xie, Ruida and Dempster, Andrew G.},
  title   = {An On-Line Deep Learning Framework for Low-Thrust Trajectory Optimisation},
  journal = {Aerospace Science and Technology}, volume = {118}, pages = {107002}, year = {2021},
  doi     = {10.1016/j.ast.2021.107002}
}

@article{Guo2023TransferTime,
  author  = {Guo, Xiang and Ren, Di and Wu, Di and Jiang, Fanghua},
  title   = {{DNN} Estimation of Low-Thrust Transfer Time: Focusing on Fast Transfers in Multi-Asteroid Rendezvous Missions},
  journal = {Acta Astronautica}, volume = {204}, pages = {518--530}, year = {2023},
  doi     = {10.1016/j.actaastro.2022.09.006}
}

@article{Ozaki2022AsteroidDNN,
  author  = {Ozaki, Naoya and others},
  title   = {Asteroid Flyby Cycler Trajectory Design Using Deep Neural Networks},
  journal = {Journal of Guidance, Control, and Dynamics}, volume = {45}, number = {8}, pages = {1496--1511}, year = {2022},
  doi     = {10.2514/1.G006487}
}

@article{Izzo2025ACTFriends,
  author  = {Izzo, Dario and M{\"a}rtens, Marcus and Beauregard, Laurent and Bannach, Max and Acciarini, Giacomo and Blazquez, Emmanuel and Hadjiivanov, Alexander and Grover, Jai and Hei{\ss}el, Gernot and Shimane, Yuri and Yam, Chit Hong},
  title   = {Asteroid Mining: {ACT}\&Friends' Results for the {GTOC12} Problem},
  journal = {Astrodynamics}, volume = {9}, number = {1}, pages = {19--40}, year = {2025},
  doi     = {10.1007/s42064-024-0204-x}
}

@inproceedings{Hennes2016FastApproximators,
  author    = {Hennes, Daniel and Izzo, Dario and Landau, Damon},
  title     = {Fast Approximators for Optimal Low-Thrust Hops Between Main Belt Asteroids},
  booktitle = {Proceedings of the 2016 IEEE Symposium Series on Computational Intelligence (SSCI)},
  year      = {2016},
  pages     = {1--7},
  publisher = {IEEE},
  doi       = {10.1109/SSCI.2016.7850107}
}

@inproceedings{Acciarini2026MIMReachability,
  author    = {Acciarini, Giacomo and Izzo, Dario and Zhang, Zhong},
  title     = {Reachability for Low-Thrust Trajectories via Maximum Initial Mass},
  booktitle = {Proceedings of the 30th International Symposium on Space Flight Dynamics},
  year      = {2026},
  pages     = {1--11},
  url       = {https://arxiv.org/abs/2605.23770}
}

@misc{Zhang2026PretrainedApproximators,
  author        = {Zhang, Zhong and Acciarini, Giacomo and Izzo, Dario and Baoyin, Hexi and Topputo, Francesco},
  title         = {Pretrained Approximators for Low-Thrust Trajectory Cost and Reachability},
  year          = {2026},
  eprint        = {2605.26790},
  archivePrefix = {arXiv},
  doi           = {10.48550/arXiv.2605.26790}
}

@article{Zhang2024MANN,
  author  = {Zhang, Zhe and Shi, Yishuai and Han, Hongwei},
  title   = {Multivariate Attention-Based Orbit Uncertainty Propagation and Orbit Determination Method for Earth--Jupiter Transfer},
  journal = {Applied Sciences},
  volume  = {14},
  number  = {10},
  pages   = {4263},
  year    = {2024},
  doi     = {10.3390/app14104263}
}

@inproceedings{Nix1994MeanVariance,
  author    = {Nix, David A. and Weigend, Andreas S.},
  title     = {Estimating the Mean and Variance of the Target Probability Distribution},
  booktitle = {Proceedings of the 1994 IEEE International Conference on Neural Networks},
  volume    = {1}, pages = {55--60}, year = {1994}, doi = {10.1109/ICNN.1994.374138}
}

@article{Wang2018Convex,
  author  = {Wang, Zhenbo and Grant, Michael J.},
  title   = {Minimum-Fuel Low-Thrust Transfers for Spacecraft: A Convex Approach},
  journal = {IEEE Transactions on Aerospace and Electronic Systems}, volume = {54}, number = {5}, pages = {2274--2290}, year = {2018},
  doi     = {10.1109/TAES.2018.2812558}
}

@inproceedings{Mao2016SCvx,
  author    = {Mao, Yuanqi and Szmuk, Michael and Acikmese, Behcet},
  title     = {Successive Convexification of Non-Convex Optimal Control Problems and Its Convergence Properties},
  booktitle = {2016 IEEE 55th Conference on Decision and Control}, pages = {3636--3641}, year = {2016},
  doi       = {10.1109/CDC.2016.7798816}
}

@article{Mao2017StateConstraints,
  author  = {Mao, Yuanqi and Dueri, Daniel and Szmuk, Michael and Acikmese, Behcet},
  title   = {Successive Convexification of Non-Convex Optimal Control Problems with State Constraints},
  journal = {IFAC-PapersOnLine}, volume = {50}, number = {1}, pages = {4063--4069}, year = {2017},
  doi     = {10.1016/j.ifacol.2017.08.789}
}

@article{Zhao2023AdaptiveSCP,
  author  = {Zhao, Jisong and Li, Jia and Li, Shuang},
  title   = {Low-Thrust Transfer Orbit Optimization Using Sequential Convex Programming and Adaptive Mesh Refinement},
  journal = {Journal of Spacecraft and Rockets}, year = {2023}, doi = {10.2514/1.A35817}
}

@article{Shirazi2018Review,
  author  = {Shirazi, Abolfazl and Ceberio, Josu and Lozano, Jose A.},
  title   = {Spacecraft Trajectory Optimization: A Review of Models, Objectives, Approaches and Solutions},
  journal = {Progress in Aerospace Sciences}, volume = {102}, pages = {76--98}, year = {2018},
  doi     = {10.1016/j.paerosci.2018.07.007}
}

\end{document}